\documentclass[twocolumn,floatfix,aps, prl]{revtex4-1}

\usepackage{graphicx}
\usepackage{mathtools}
\usepackage{amsmath}
\usepackage{amssymb}
\usepackage{bm}
\usepackage{color}
\usepackage{comment}
\usepackage[FIGTOPCAP]{subfigure}
\usepackage[]{footmisc}
\usepackage{lineno}
\usepackage{dsfont}
\usepackage{soul,xcolor}
\usepackage[pdfborder={0 0 0}]{hyperref}

\newcommand{\vv}{{\bm v}}
\newcommand{\vj}{{\bm j}}
\newcommand{\vx}{{\bm r}}
\newcommand{\vn}{{\boldsymbol \nabla }}
\newcommand{\vk}{{\boldsymbol \kappa}}

\begin{document}

\title{Parabolic Hall effect due to Co-Propagating Surface Modes}

\author{M.\ Breitkreiz}
\email{breitkr@physik.fu-berlin.de}
\affiliation{Dahlem Center for Complex Quantum Systems and Fachbereich Physik, Freie Universit\" at Berlin, 14195 Berlin, Germany}

\date{November 2019}

\begin{abstract}
Real-space separations of counter-moving states 
to opposite surfaces or edges are associated with 
different types of Hall effects, such as 
the quantum-, spin-, or the anomalous Hall effect. 
Some systems provide
the possibility to separate a fraction of countermovers in a 
completely different fashion: Surface states propagating
all in the same direction, balanced by 
counter-moving bulk states, realized,
e.g., in Weyl metals with intrinsically or
extrinsically broken inversion and time-reversal 
symmetries.
In this work we show that these co-propagating surface modes
are associated with a novel Hall effect --- a parabolic potential profile in the direction perpendicular to and in its magnitude linear
in the applied field. 
While in 2D systems the parabolic potential profile is measurable
directly, in 3D the resulting voltage between bulk and surface is 
measurable in the geometry of a hollow cylinder. Moreover, 
the parabolic Hall effect leads to characteristic signatures in the longitudinal conductivity. 
\end{abstract}

\maketitle

\textit{Introduction}---The condensed-matter
realization of Weyl fermions 
\cite{Xu2015a, Xu2015b, Lv2015, Borisenko2014, Neupane2014, Liu2014a, Xiong2015a, Armitage2017, Yan2017}
has attracted very much interest in the past years, due in large part to the realization of chiral Landau levels, moving parallel or 
antiparallel to the magnetic field, depending on the 
Weyl-fermion chirality \cite{Nielsen1983}. 
In a crystal, the two chiralities appear
pairwise, separated in momentum space, 
which complicates an
identification of Weyl-specific transport phenomena such as 
the chiral magnetic effect \cite{Burkov2017,Burkov2017a, Nandy2017, Reis2016}.

\begin{figure}[t]
\includegraphics[width=\columnwidth]{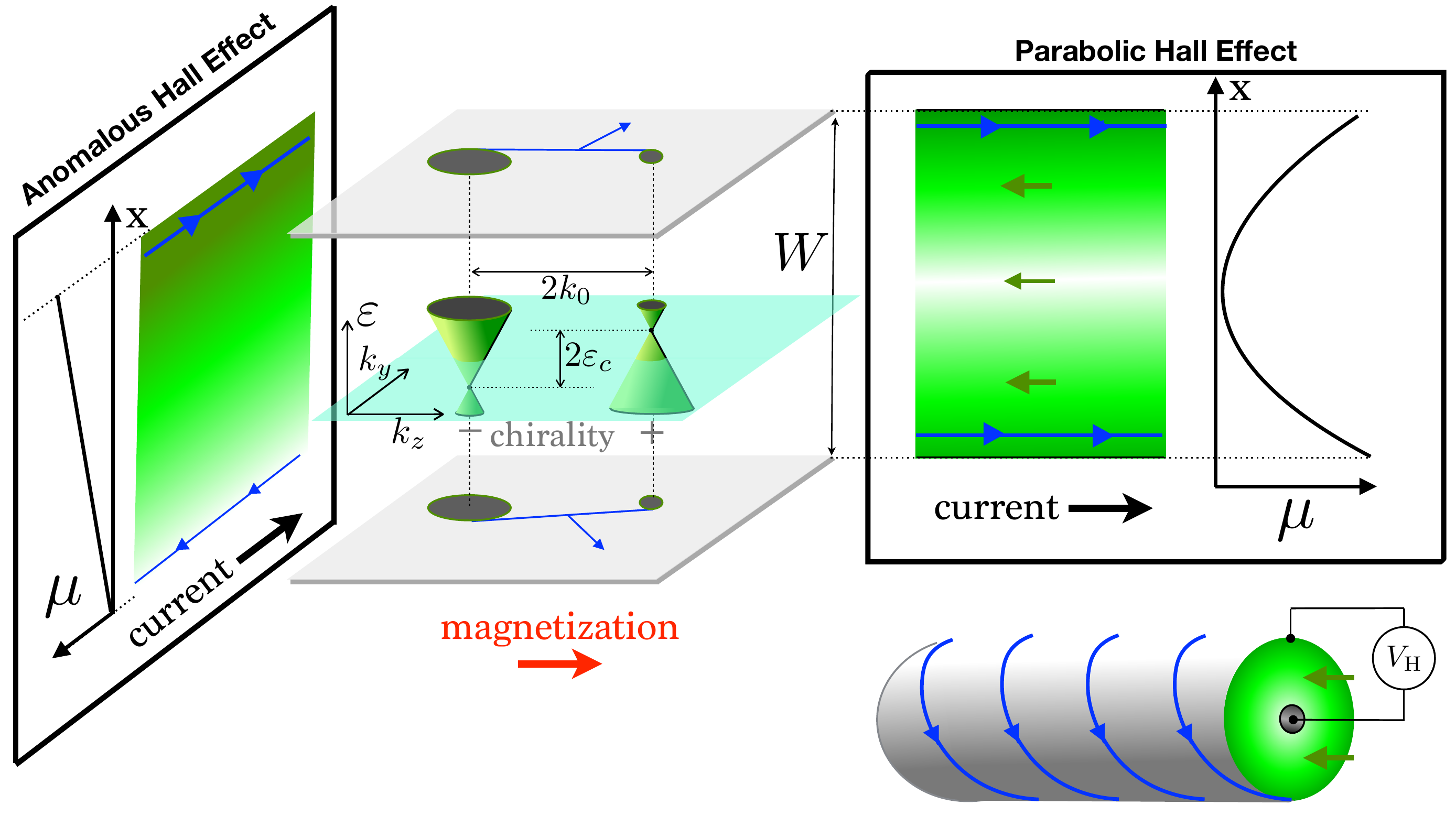}
\caption{Weyl-metal slab with two Weyl cones separated
in energy and in momentum, connected by Fermi arcs
at the surfaces, illustrated in a mixed momentum/real space.  
Anomalous Hall effect --- a linear drop of the chemical potential $\mu$ between the surfaces --- occurs if current is led
in the plane of counter-propagating surface modes. 
The parabolic Hall 
effect occurs when current is led in the plane of co-propagating
surface modes, leading to a quadratic
spatial dependence of $\mu$, measurable
in the geometry of a hollow cylinder between the inner and outer surfaces.}
\label{fig1a}
\end{figure}

Tendentiously it is more promising when chiral 
states are separated not (or not only) in momentum
but in real space --- a situation well known from the 
separation of 
countermovers to opposite surfaces in 
topological insulators \cite{Qi2011}. Here the favorable situation
associated with the real-space separation is 
the clear signature in form of a quantum (spin) Hall effect
 \cite{VonKlitzing1986, Konig2007}. 
 A noteworthy equivalent 
in the field of Weyl metals is the anomalous Hall 
effect (AHE) \cite{Burkov2011, Burkov2014} --- a voltage drop in the direction perpendicular to
both, the direction of the current flow and the intrinsic
magnetization, in the absence of an external magnetic 
field \cite{Suzuki2016, Liu2017c, Li2019}.
The mechanism of the AHE in Weyl metals
can indeed be understood in terms of chiral surface states ---
a pairwise connection of 
Weyl Fermi surfaces of opposite chirality by two Fermi arcs
\cite{Balents2011},
localized at opposite surfaces and
moving in opposite directions, which intuitively explains
the AHE as the contribution of Fermi arcs 
in case of a potential difference between the surfaces \cite{Armitage2017, Breitkreiz2019}, see Fig.\ \ref{fig1a}.
Remarkably, the presence of a finite density of diffusive 
bulk states does not obscure the chiral-surface-states driven AHE.

In this work we focus on a different and much less explored separation of countermovers in real space: 
forwardmovers  homogeneously
distributed in the bulk and backmovers localized at the surface.
This  
can be realized in
 2D systems \cite{Colomes2018} (edge states have been called ``antichiral'' 
in this case), including transition-metal dichalcogenide monolayers \cite{Colomes2018}, and exciton-polariton 
systems \cite{Mandal2019}, and in 
 3D Weyl metals \cite{Baireuther2016, Pikulin2016}. 
 Comparing to the well-studied case of counter-propagating
 surface modes the question
 arises, whether co-propagating surface modes can also be associated
 with a characteristic Hall effect.

We show that the answer is positive ---
 co-propagating surface modes give rise to a novel Hall response, characterized
by a \emph{quadratic}   
 spatial dependence of the chemical potential $\mu$, transverse to the applied field.
The general mechanism can be understood as follows.
In the bulk, the current density of chiral charge carriers (the fraction of bulk charges compensating the current of co-propagating surface states) changes
proportional to the local chemical potential, $\delta\vj_c\propto\delta\mu$. Together with the coexisting diffusive bulk 
charges flowing according to $\vj_n \propto \vn \mu$, the total current must be divergence free in the steady state,
hence $\vn^2\mu \propto -\hat{\bm{j}_c}\cdot\vn \mu$, where
$ \hat{\bm{j}_c}$ is the direction of the chiral bulk current.
If now a homogeneous electric field $\bm{E}=-\vn \mu$ is applied along  $\hat{\bm{j}_c}$, the chemical potential 
assumes a quadratic spatial dependence  in the direction 
perpendicular to the applied field. This is what we call
the parabolic Hall effect (PHE). Note that this effect is 
still linear in the driving field and thus distinct from 
 effects called ``non-linear Hall effect''  \cite{Sodemann2015}.
 
 In the following we explore the effect in detail. We 
focus on a minimal 3D model of a Weyl metal, Fig.\ \ref{fig1a}, which, while
 allowing direct conclusion on the simpler case of 2D, requires additional calculations to show how the resulting Hall voltage can be measured in 3D systems.

\emph{Model}---Measuring energy in units of $\hbar v$, where $v$ is the Fermi 
velocity, and length in units of the lattice constant, the 
  Hamiltonian we consider reads
\begin{equation}
H=- \sigma_x i\partial_x + k_y\sigma_y  + m(k_z)\sigma_z+ \varepsilon_c \eta(k_z), \label{ham}
\end{equation}
where $\sigma_i$ are spin Pauli matrices, $m(k_z)=(k_z^2-k_0^2)/2k_0$,  and $\eta(k_z) = \tanh( 2 k_z/k_0)$, featuring two Weyl nodes with 
chirality $\pm$ at momentum  $k_{x,y}=0, \;k_z=\pm k_0$ and energy $\varepsilon\approx \pm \varepsilon_c$.
 
We focus on the case of two well-separated Weyl cones 
with vanishing corrections to the linear dispersion and 
a constant velocity $v$ at the Fermi level $\varepsilon_F$, 
hence 
$\varepsilon_c,|\varepsilon_F| \ll k_0$. The explicit 
form of $m(k_z)$ and $\eta(k_z)$ is unimportant as long as these 
requirements are fulfilled.

\begin{figure}[b]
\includegraphics[width=0.9\columnwidth]{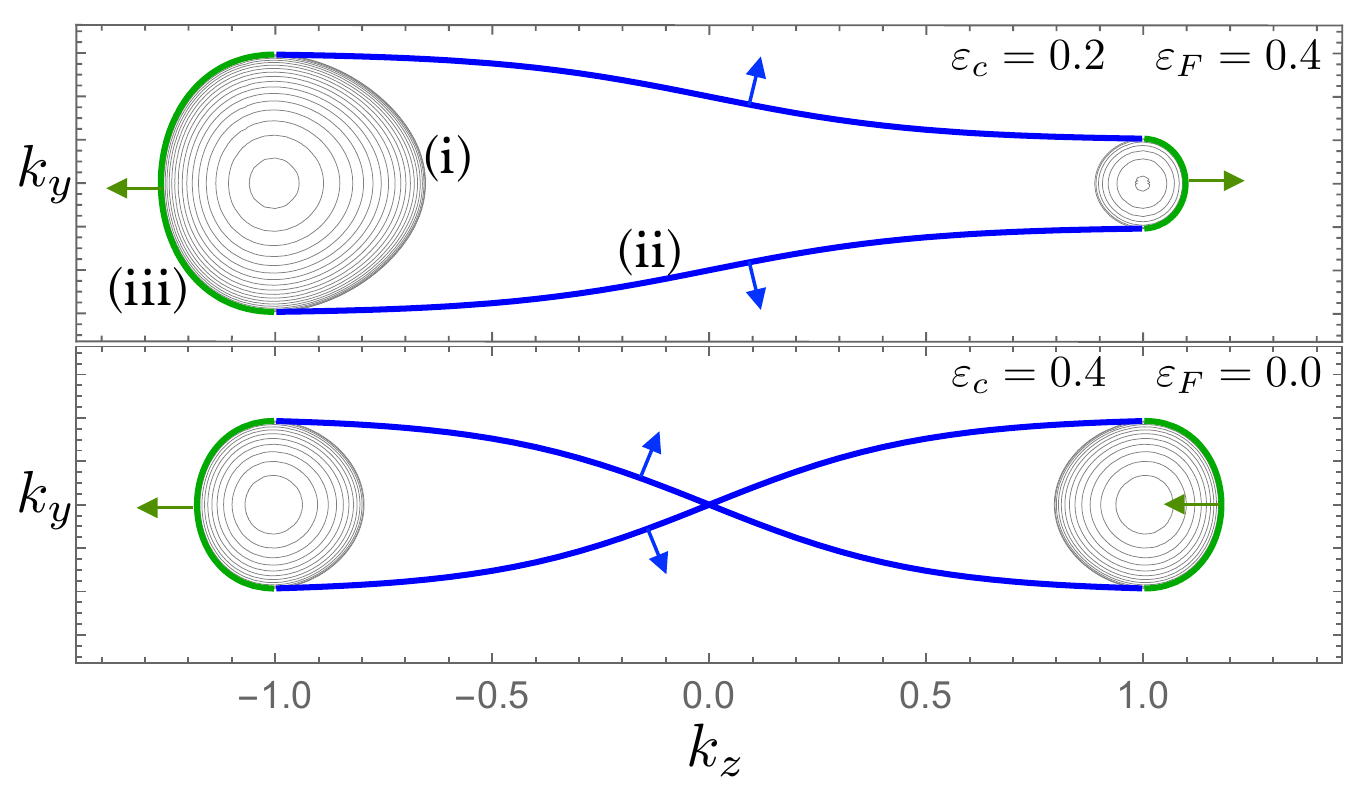}
\caption{Fermi-level states for the 
Weyl metal of width $W=100$, $k_0=1$, and two 
different combinations of $\varepsilon_F$ and 
$\varepsilon_c$. The 
gray, blue, and green states correspond to (i) normal
bulk state, (ii) surface states, and (iii)  chiral bulk states, respectively.
Velocity directions are indicated by arrows.}
\label{fig1}
\end{figure}

Considering a slab of width $W$ the quantum numbers  
are $\vk = (q, k_y, k_z) $, where $q$ is 
the solution of 
\begin{align}
&\frac{m(k_z)}{q}\tan (W\,q)+1=0
\end{align}
coming from the boundary condition of the slab \cite{Bovenzi2017a}.
Throughout this work we assume $W\gg 1$, in which 
case the solution for $q$ can be divided into three groups,
(i)  the group of bulk states,
given by the quasi continuous set $q=(n+0.5)\pi/W$, $n=0,1,2,\dots$,
(ii) surface states with the imaginary
(and hence exponentially decaying) 
solution $q=i \, m(k_z)$ for  $ m(k_z)<0$, and
(iii)  chiral bulk states with the 
solution  $q= 0$ for $ m(k_z) >0$.
The dispersion reads
\begin{equation}
\varepsilon_\vk = \varepsilon_c\eta(k_z) \pm \sqrt{q^2 + k_y^2 +[m(k_z)]^2}
\end{equation}
and the equi-energy contours are illustrated in 
Fig.\ \ref{fig1}. Note that surface states (ii) and 
chiral bulk states (iii) merge at $m(k_z)=0$ building a
single closed contour. 

Since the $x$ dependence of the wavefunctions is 
given by $\exp(i\, q\,  x)$, the finite penetration depth of surface states is $1/\mathrm{Im}\, q$.
The velocity 
of a wave packet at state $\vk$ can thus be written as ${\bm v}_\vk = 
\mathrm{Re}[(\partial_q,\partial_{k_y}, \partial_{k_z})\varepsilon_\vk]$ and the free spectral function
in the center-of-mass coordinate $x$
and the  $W\gg 1$ limit reads 
\begin{align}
A(\vk,x,\omega) &=  2\pi \delta(\omega-\varepsilon_\vk)\rho(\vk,x),
 \label{A}\\
\rho(\vk,x) &= 
 \begin{cases} 
1 &  \mathrm{Im}\, q = 0 \\   
2Wm(k_z)\; e^{\pm 2m(k_z) (x\mp W/2)} & \mathrm{Im}\, q \neq 0
\end{cases} 
\label{rho}
\end{align}
where $\mathrm{Im}\, q=0$ and
 $\mathrm{Im}\, q\neq 0$ distinguishes bulk and surface 
states, and $\pm$ corresponds to surface states at $x=\pm W/2$. Despite the divergence of 
penetration at $m(k_z)=0$, the surface-state  
spectral weight averaged over all states at the surface is strongly 
localized, the characteristic length scale being $1/k_0 \sim 1$. 
Since all 
other length scales will be considered to be much larger, 
we approximate the $ \mathrm{Im}\, q \neq 0$ case in
\eqref{rho} as $\rho(\vk,x) \approx W\delta(x\mp W/2)$
in the following.

The density of bulk states 
at the Fermi level of the cone $\pm$ and the total 
bulk density $n_n = \sum_\pm n_{n\pm}$, read, respectively,
\begin{equation}
n_{n\pm} = \frac{(\varepsilon_F\mp \varepsilon_c)^2}{\pi v h},
\;\;\;\;\;
n_n = 2\frac{\varepsilon_F^2+\varepsilon_c^2}{\pi v h}.
\end{equation}
We also define the density of chiral bulk states and the 
2D density of surface states of a single surface, 
\begin{equation}
n_{c\pm} = \frac{|\varepsilon_F\mp \varepsilon_c|
}{2 v h W},\;\;\;\;\;\;
n_s = \frac{k_0}{\pi v h},
\end{equation}
respectively. In accord with $W\gg 1$, we assume that $n_n$ is much larger than $n_{c\pm}$
and $n_s/W$.

\emph{Parabolic Hall effect}---To explore the transport behavior in linear response, we aim to find a solution for
a state-dependent
deviation of the chemical potential from the Fermi energy, 
$\mu(\vk)$, with an arbitrary spatial profile along the 
$x$ direction, given the boundary condition of a homogeneous
force field applied in the $z$ direction, $\partial_z\mu(\vk) = E$.
Furthermore we assume
 elastic scattering from a weak disorder potential.
To focus on
qualitative features, we take the 
scattering amplitudes 
different only between Fermi-level states of the different types $i \in[ n+,c+,s+,n-,c-,s-]$.  
Using the Quantum Boltzmann formalism \cite{Mahan2000} and  
employing the standard semiclassical approximation scheme, see Supplemental Material for details
\cite{*[{See Supplemental Material for details of the 
derivation and solution of transport equations for the
slab and the hollow-cylinder geometry}] [{}] dummy2}, 
we obtain
 \begin{subequations}
\begin{align}
\vn \cdot(\vj_{n\pm}+\vj_{c\pm}) = & \pm  \gamma_{n- n+}(\mu_{n-}-\mu_{n+}) \nonumber\\
& \;\;\;\;\;\;\;+\gamma_{sn\pm}s(x) (\mu_s-\mu_{n\pm}), \label{eq1}\\
s(x)\; \vn \cdot \vj_{s\pm} =&\;  s(x)\big[
\gamma_{sn+}(\mu_{n+}-\mu_{s\pm}) \nonumber\\
&\;\;\;\;\;\;\;\; +\gamma_{sn-}(\mu_{n-}-\mu_{s\pm})\big], \\
\vj_{n\pm} = & -n_{n\pm}D\, \vn\mu_{n\pm}, 
 \label{jnpm}
\end{align}
\label{eqs}
\end{subequations}
where $\gamma_{ij} =\Gamma_{ij}n_i n_j$ and 
$\Gamma_{ij}$ is the scattering rate, 
$\mu_i = \langle \mu(\vk)\rangle_i$ is 
the local chemical potential averaged over the 
Fermi-level states $i$,
$\vj_i = n_i\langle {\bm v}_\vk\mu(\vk)\rangle_i$  is the non-equilibrium current-density contribution of the states $i$,
$D$ is the bulk diffusion constant,  and 
$s(x) = \sum_\pm \delta(x\pm W/2) $. 

To linear order in $E$ and using translation invariance
in the $y$ direction, the divergence of the 
chiral-bulk and the surface particle currents simplify to
\begin{align}
\vn\cdot  \vj_{c\pm} &= 
\pm\frac{\varepsilon_F\mp \varepsilon_c}{\pi h W}\, E,
\;\;\;\;\;\; 
\vn\cdot \vj_{s\pm} = \frac{\varepsilon_c}{\pi h} E. \label{jcpm} 
\end{align}
We rewrite Eqs.\ \eqref{eqs} by considering the differential 
Eqs.\ \eqref{eq1} away from the boundary ($s(x)=0$),
together with \eqref{jnpm} and \eqref{jcpm},
\begin{equation}
\pm\frac{\varepsilon_F\mp \varepsilon_c}{\pi h W} E
-n_{n\pm}D \partial^2_x \mu_{n\pm} = \pm
\gamma_{n+n-}
(\mu_{n-}-\mu_{n+}). \label{eqb}
\end{equation}
This is supplemented with boundary conditions, given by integration 
of \eqref{eqs} over an infinitesimal distance at both boundaries and assuming vanishing current through the boundary,
\begin{subequations}
\begin{align}
j^x_{n+}(\pm W/2) =& \mp
\gamma_{sn+}(\mu_{s\pm}-\mu_{n+}(\pm W/2)), \\
j^x_{n-}(\pm W/2) =& \mp
\gamma_{sn-} (\mu_{s\pm}-\mu_{n-}(\pm W/2)), \\
 \frac{\varepsilon_c}{\pi h} E =&   
\gamma_{sn+} (\mu_{n+}(\pm W/2)-\mu_{s\pm}) \nonumber\\
&+\gamma_{sn-} (\mu_{n-}(\pm W/2)-\mu_{s\pm}).
\end{align}
\label{bcs}
\end{subequations}
Assuming that an 
external contact would couple to all bulk states with 
equal probability it would probe the 
averaged chemical potential
\begin{equation}
\mu_{n} = \frac{n_{n+}\mu_{n+}+ n_{n-}\mu_{n-}}{n_n},
\end{equation}
for which Eqs.\ \eqref{eqb} and \eqref{bcs} readily provide the solution
\begin{equation}
\mu_n = -\frac{n_s v}{n_nD}\, \frac{\varepsilon_c}{k_0} \frac{x^2}{W}\, E+z \, E.
\label{phr2}
\end{equation}
Sticking to the interpretation that 
$ E$ is an applied field in the $z$ direction, the response
lies in the first term in \eqref{phr2}, which 
exhibits the PHE --- a 
quadratic spatial dependence on the transverse
coordinate $x$ and a linear dependence on the magnitude
of the applied field $E$. The roles of applied and induced 
fields are of course interchangeable so that a finite 
magnitude of the transverse potential would induce a finite
longitudinal field $E$ and with that a longitudinal current
(which will be calculated below). 

\begin{figure}[]
\includegraphics[width=0.7\columnwidth]{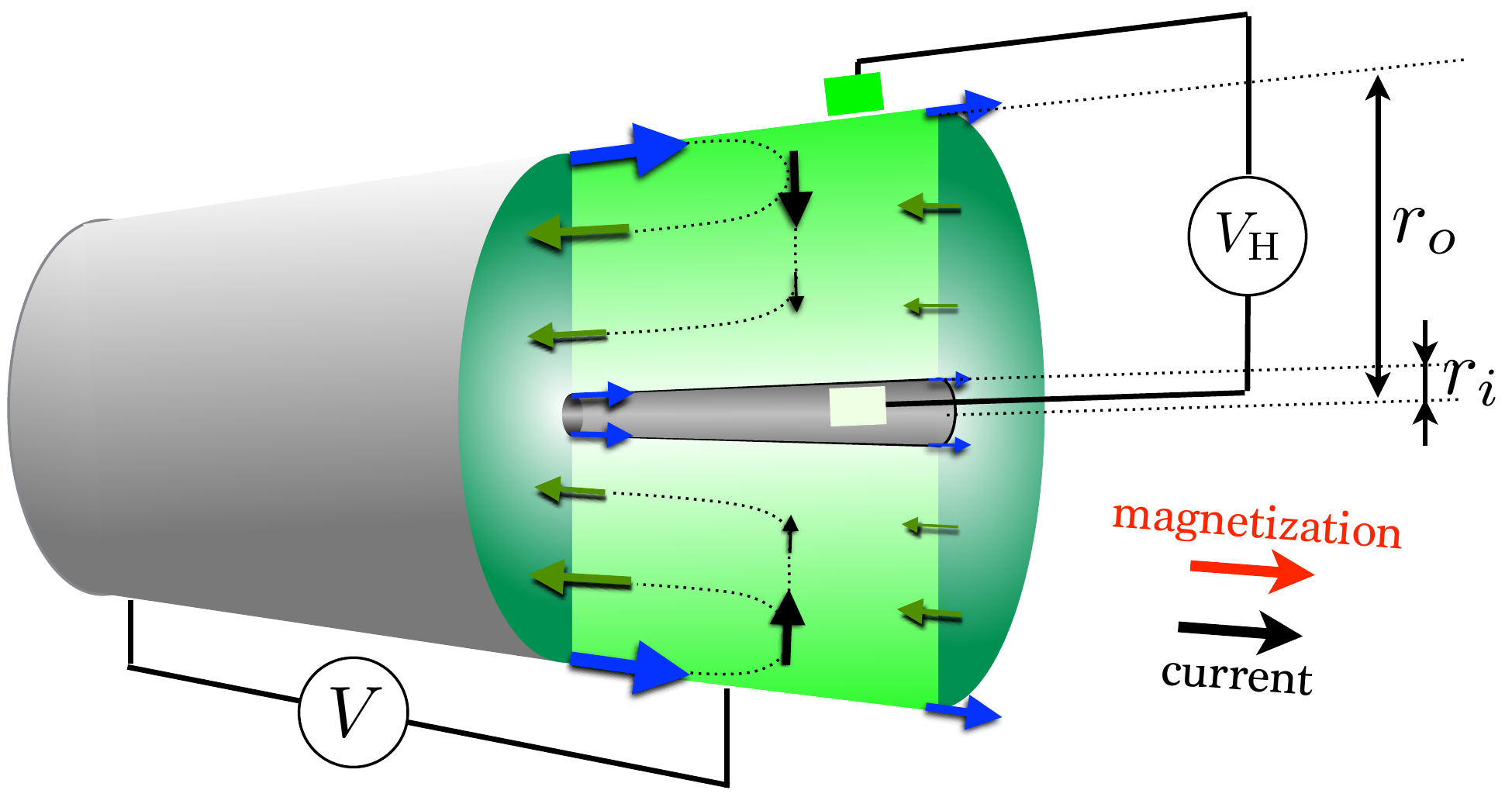}
\caption{Weyl metal in the geometry of a hollow
 cylinder with a small inner radius. 
When current is led along the cylinder and the magnetization, the PHE voltage is induced between the  
inner and the outer surface. Arrows indicate local
current flows of chiral bulk (green), surface (blue), 
and diffusive bulk (black) particles.}
\label{fig2}
\end{figure}

\emph{Measuring the Hall voltage}---In contrast to the ordinary Hall effect where the potential varies linearly and is fully
characterized by a voltage between opposite surfaces, 
the PHE voltage occurs between one surface and
 the bulk and
 varies quadratically with the distance. While in a 2D system
  the parabolic voltage profile can be measured directly,
in a 3D system a contact inside the sample necessarily introduces an inner surface so that the actual geometry for such a measurement is (in its most simple realization) 
that of a hollow cylinder, see Fig.\ \ref{fig2},
with an inner 
radius $r_i$ and an outer radius $r_o$. 
A straightforward modification of the above formalism 
\cite{dummy2}
[essentially consisting in the 
replacements $W\to r_o-r_i$, $x\to r$,
$\partial_x j^x_n \to (\partial_r+1/r)j^r_n$]
leads to the solution
\begin{equation}
\mu_n(r) = -\frac{n_s v}{n_nD} \frac{\varepsilon_c}{k_0}
\bigg( \frac{r^2/2}{r_o-r_i} 
-\frac{r_or_i}{r_o-r_i}\ln\frac{r}{r_i}\bigg)E +zE.
\label{cyl}
\end{equation}
We quantify the voltage between the inner and the outer surfaces via the resulting Hall angle $
\theta_\mathrm{PHE} \equiv [\mu_n(r_o)-
\mu_n(r_i)]/(r_o-r_i)E$.
In the limit $r_o\gg r_i$ we obtain
\begin{equation}
\theta_\mathrm{PHE} = -\frac{n_s v}{n_nD}\frac{\varepsilon_c}{2 k_0}.\label{phe}
\end{equation}
Note that the first factor,
 $n_s v / n_nD\equiv \theta $, 
is the Hall angle of the AHE \cite{Burkov2014, Breitkreiz2019}, which is thus related to 
$\theta_\mathrm{PHE} $ by the ratio of 
energy- vs.\ momentum separation 
of the Weyl nodes 
(energy in units of $\hbar v$).

\emph{Conductivity}---We first note that general symmetry considerations
\cite{Kleiner1966, Seemann2015} fix the form 
of the infinite-system conductivity tensor of 
our system to
\begin{equation}
\sigma = \begin{pmatrix}
\sigma^{xx} & \sigma^{xy} & 0 \\
-\sigma^{xy} & \sigma^{xx} & 0\\
0&0&\sigma^{zz}
\end{pmatrix},
\end{equation}
which follows from rotation symmetry around
$z$ and symmetry with respect to
time reversal combined with $C_2$ rotation around an
 $x$-$y$-plane axis, corresponding to the Laue
group $\infty 2'$  \cite{Kleiner1966}. As discussed above,  the model exhibits AHE in the $x$-$y$ plane,
with a Hall angle $\sigma^{xy}/\sigma^{xx} = \theta $, 
related to the
intrinsic magnetization in the $z$ direction. 
The PHE is instead found in the plane parallel to $z$ but
it evidently does not manifest itself in a finite $\sigma^{xz}$ or $\sigma^{yz}$,
which is in agreement with our result that 
the potential difference in the  $x$ direction
 between outer surfaces vanishes.

In the following we show that the PHE still manifests itself
in the infinite-system conductivity 
 --- it gives rise to an anomalous term in $\sigma^{zz}$, which is 
size-dependent but finite in the infinite-system limit.
The current contribution of normal bulk 
 states in response to the field $E$ 
 is obtained from \eqref{jnpm} as
 $j^z_n = -n_nD\, E$. This would be the only 
 contribution to $\sigma^{zz}$ that one would obtain for an 
 infinite system
 based on the Drude formula; 
 we denote it as $\sigma_0^{zz} \equiv j^z_n/E$. 
The additional contributions of chiral bulk states and surface 
states, neglecting corrections of order $1/W$, 
can be written as 
 $j^z_{c\pm}= \pm (\varepsilon_F\mp\varepsilon_c)\,\mu_{n\pm}
 /\pi h W$ and
 $j^z_{s\pm}=\delta(x\mp W/2)\,\mu_{s\pm} \varepsilon_c/\pi h $, respectively. Inserting $\mu_i$ as solutions of \eqref{eqb} 
 and \eqref{bcs}, the full conductivity is given by 
 $\sigma^{zz}  =  (j^z_n+ \bar{j}^z_c+\bar{j}^z_s)/E$, where the bar denotes  averaging over $x$. In terms of $\sigma^{zz}_0$
 we obtain
\begin{align}
\frac{\sigma^{zz}}{\sigma^{zz}_0}=&1+\frac{4}{3}\theta^2_\mathrm{PHE} \bigg(1+\frac{6l_{ns}}{W\theta}\bigg)
\bigg(1+\frac{\varepsilon_F^2}{\varepsilon_c^2}\xi\bigg),
\label{res}\\
&\xi = \frac{\frac{2 l_c}{W}-\frac{\big(\frac{2 l_c}{W}\big)^2}{\frac{ l_{ns}}{\theta l_c}+\coth\frac{W}{2l_c}}}{\frac{W}{6 l_c}+\frac{l_{ns}}{\theta l_c}}
=
\begin{cases} 
0 & l_c \ll W \\
1 & l_c \gg W,\;  l_{ns}/\theta,
\end{cases}
\end{align}
where $l_{ns} = v/\Gamma_{sn}n_n$ is the relaxation length of surface states  and 
 $l_c =\sqrt{D/\Gamma_{n+n-} n_n}$ is the internode
 relaxation length \cite{Parameswaran2014}. 
In the infinite-system limit we obtain 
$\sigma^{zz} = (1+4\theta_\mathrm{PHE}^2/3)\sigma^{zz}_0$, demonstrating a remarkable deviation from the Drude behavior.  We expect that the same 
correction can be derived based purely 
on the bulk Hamiltonian, e.g., using Kubo formalism. We leave this for future work, noting that similarly exchangeable derivations from an infinite- and a finite-system perspective 
have been demonstrated for the AHE in \cite{Burkov2014} and \cite{Breitkreiz2019}.

 The finite-size correction in the first brackets is similar to the  finite-size correction of the AHE in \cite{Breitkreiz2019} and is due to the vanishing dissipation of surface states
 if their scattering length becomes large 
 compared to the width $W$. The second 
 brackets correspond to a finite-size correction which 
 occurs if the system is not electron-hole compensated
 ($\varepsilon_F\neq 0$), in which case the applied 
field induces an occupation disbalance between the
Weyl nodes leading to a prolongated relaxation time. 
The resistivity decrease saturates even if 
$l_c \to \infty$ because internode relaxation  
also happens indirectly via surface states.

\emph{Discussion}---Our calculations have shown that 
co-propagating surface modes and the related
counter-propagating  bulk states (which we here call chiral) 
give rise to a parabolic
transverse potential profile. We made the realistic assumption
that the density of coexisting
 normal bulk states (spatially not separated countermovers) 
 is finite and hence
 $\sim W$  times larger 
than the number of surface states, where 
$W$ is the width in units of the lattice constant. Nevertheless, 
the spatial separation of surface and chiral bulk states 
by $\sim W$ compensates this so that the PHE 
survives the limit $W \to \infty $. 
Besides the Hall voltage, we have identified an anomalous term in the longitudinal conductivity, which can be interpreted as the 
precursor of the anomalously large conductance in the quantum regime of localized normal bulk states 
\cite{Colomes2018, Behrends2019}. 

We exemplified the parabolic Hall effect on a model for a
Weyl metal with intrinsically 
broken inversion and time-reversal symmetries,
which shows a homogeneous 
chiral charge density in the bulk
\cite{*[{More generally, a
non-linear potential is produced by 
chiral states acting as 
particle sources or drains when  a driving 
field $E_\parallel$ is aligned with their motion, 
$
\vn^2 \mu =  E_\parallel/\lambda,
$
where the characteristic length $\lambda$ given
by the relative densities and mobilities of the chiral and 
diffusive particles might be spatially inhomogeneous}] [{}] dummy}.
Several ways have been proposed to
induce (or change) the chiral charge density in Weyl metals externally.
For example, internode charge pumping via parallel 
 electric and magnetic fields can effectively shift
 the energies of the Weyl nodes as
\begin{equation}
\varepsilon_c\to \varepsilon_c + \frac{e^2}{h^2}\frac{\tau}{n_n} E B, 
\end{equation} 
where $\tau$ is the internode relaxation time, which is 
assumed to be much larger than intranode relaxation
\cite{Burkov2017}. Another interesting possibility 
to induce and control the PHE 
is the application of a 
strain-induced pseudomagnetic field \cite{Pikulin2016, Grushin2016, Behrends2019},
which allows to change the density of chiral states even 
without an electric field and independent of the
internode relaxation.  

Furthermore, the derived unconventional
dependence of the conductivity on the system size and 
the scattering amplitudes as a consequence of the 
PHE is an interesting starting and reference point for 
investigations of the conductivity with regard to temperature dependence
in case of phonon-mediated scattering \cite{Pereira2019}, 
or the scaling 
behaviour with the system size \cite{Zhang2019}.
It is worth noting that in a time-reversal invariant Weyl metal, the mechanism of a reduced and size-dependent 
resistivity  due to a doubling of the AHE 
\cite{Breitkreiz2019}, also 
applies to the PHE. The 
PHE voltage would vanish but the associated suppression of the resistivity
would remain when the coupling between the time-reversed
states is sufficiently weak. 
The suppression would set in when the width $W$ becomes
comparable to the characteristic scattering length quantifying the coupling of time-reversed states.

\textit{Acknowledgments}. The author would like to thank 
Piet W. Brouwer, Tobias Meng, and Rodrigo G. Pereira for valuable discussions. 
This research was supported by the Grant
No. 18688556 of the Deutsche Forschungsgemeinschaft
(DFG, German Research Foundation).

\bibliography{library}	

\begin{thebibliography}{40}%
\makeatletter
\providecommand \@ifxundefined [1]{%
 \@ifx{#1\undefined}
}%
\providecommand \@ifnum [1]{%
 \ifnum #1\expandafter \@firstoftwo
 \else \expandafter \@secondoftwo
 \fi
}%
\providecommand \@ifx [1]{%
 \ifx #1\expandafter \@firstoftwo
 \else \expandafter \@secondoftwo
 \fi
}%
\providecommand \natexlab [1]{#1}%
\providecommand \enquote  [1]{``#1''}%
\providecommand \bibnamefont  [1]{#1}%
\providecommand \bibfnamefont [1]{#1}%
\providecommand \citenamefont [1]{#1}%
\providecommand \href@noop [0]{\@secondoftwo}%
\providecommand \href [0]{\begingroup \@sanitize@url \@href}%
\providecommand \@href[1]{\@@startlink{#1}\@@href}%
\providecommand \@@href[1]{\endgroup#1\@@endlink}%
\providecommand \@sanitize@url [0]{\catcode `\\12\catcode `\$12\catcode
  `\&12\catcode `\#12\catcode `\^12\catcode `\_12\catcode `\%12\relax}%
\providecommand \@@startlink[1]{}%
\providecommand \@@endlink[0]{}%
\providecommand \url  [0]{\begingroup\@sanitize@url \@url }%
\providecommand \@url [1]{\endgroup\@href {#1}{\urlprefix }}%
\providecommand \urlprefix  [0]{URL }%
\providecommand \Eprint [0]{\href }%
\providecommand \doibase [0]{http://dx.doi.org/}%
\providecommand \selectlanguage [0]{\@gobble}%
\providecommand \bibinfo  [0]{\@secondoftwo}%
\providecommand \bibfield  [0]{\@secondoftwo}%
\providecommand \translation [1]{[#1]}%
\providecommand \BibitemOpen [0]{}%
\providecommand \bibitemStop [0]{}%
\providecommand \bibitemNoStop [0]{.\EOS\space}%
\providecommand \EOS [0]{\spacefactor3000\relax}%
\providecommand \BibitemShut  [1]{\csname bibitem#1\endcsname}%
\let\auto@bib@innerbib\@empty
\bibitem [{\citenamefont {Xu}\ \emph {et~al.}(2015{\natexlab{a}})\citenamefont
  {Xu}, \citenamefont {Alidoust}, \citenamefont {Belopolski}, \citenamefont
  {Yuan}, \citenamefont {Bian}, \citenamefont {Chang}, \citenamefont {Zheng},
  \citenamefont {Strocov}, \citenamefont {Sanchez}, \citenamefont {Chang},
  \citenamefont {Zhang}, \citenamefont {Mou}, \citenamefont {Wu}, \citenamefont
  {Huang}, \citenamefont {Lee}, \citenamefont {Huang}, \citenamefont {Wang},
  \citenamefont {Bansil}, \citenamefont {Jeng}, \citenamefont {Neupert},
  \citenamefont {Kaminski}, \citenamefont {Lin}, \citenamefont {Jia},\ and\
  \citenamefont {{Zahid Hasan}}}]{Xu2015a}%
  \BibitemOpen
  \bibfield  {author} {\bibinfo {author} {\bibfnamefont {S.-Y.}\ \bibnamefont
  {Xu}}, \bibinfo {author} {\bibfnamefont {N.}~\bibnamefont {Alidoust}},
  \bibinfo {author} {\bibfnamefont {I.}~\bibnamefont {Belopolski}}, \bibinfo
  {author} {\bibfnamefont {Z.}~\bibnamefont {Yuan}}, \bibinfo {author}
  {\bibfnamefont {G.}~\bibnamefont {Bian}}, \bibinfo {author} {\bibfnamefont
  {T.-R.}\ \bibnamefont {Chang}}, \bibinfo {author} {\bibfnamefont
  {H.}~\bibnamefont {Zheng}}, \bibinfo {author} {\bibfnamefont {V.~N.}\
  \bibnamefont {Strocov}}, \bibinfo {author} {\bibfnamefont {D.~S.}\
  \bibnamefont {Sanchez}}, \bibinfo {author} {\bibfnamefont {G.}~\bibnamefont
  {Chang}}, \bibinfo {author} {\bibfnamefont {C.}~\bibnamefont {Zhang}},
  \bibinfo {author} {\bibfnamefont {D.}~\bibnamefont {Mou}}, \bibinfo {author}
  {\bibfnamefont {Y.}~\bibnamefont {Wu}}, \bibinfo {author} {\bibfnamefont
  {L.}~\bibnamefont {Huang}}, \bibinfo {author} {\bibfnamefont {C.-C.}\
  \bibnamefont {Lee}}, \bibinfo {author} {\bibfnamefont {S.-M.}\ \bibnamefont
  {Huang}}, \bibinfo {author} {\bibfnamefont {B.}~\bibnamefont {Wang}},
  \bibinfo {author} {\bibfnamefont {A.}~\bibnamefont {Bansil}}, \bibinfo
  {author} {\bibfnamefont {H.-T.}\ \bibnamefont {Jeng}}, \bibinfo {author}
  {\bibfnamefont {T.}~\bibnamefont {Neupert}}, \bibinfo {author} {\bibfnamefont
  {A.}~\bibnamefont {Kaminski}}, \bibinfo {author} {\bibfnamefont
  {H.}~\bibnamefont {Lin}}, \bibinfo {author} {\bibfnamefont {S.}~\bibnamefont
  {Jia}}, \ and\ \bibinfo {author} {\bibfnamefont {M.}~\bibnamefont {{Zahid
  Hasan}}},\ }\href {\doibase 10.1038/nphys3437} {\bibfield  {journal}
  {\bibinfo  {journal} {Science}\ }\textbf {\bibinfo {volume} {349}},\ \bibinfo
  {pages} {613} (\bibinfo {year} {2015}{\natexlab{a}})}\BibitemShut {NoStop}%
\bibitem [{\citenamefont {Xu}\ \emph {et~al.}(2015{\natexlab{b}})\citenamefont
  {Xu}, \citenamefont {Alidoust}, \citenamefont {Belopolski}, \citenamefont
  {Yuan}, \citenamefont {Bian}, \citenamefont {Chang}, \citenamefont {Zheng},
  \citenamefont {Strocov}, \citenamefont {Sanchez}, \citenamefont {Chang},
  \citenamefont {Zhang}, \citenamefont {Mou}, \citenamefont {Wu}, \citenamefont
  {Huang}, \citenamefont {Lee}, \citenamefont {Huang}, \citenamefont {Wang},
  \citenamefont {Bansil}, \citenamefont {Jeng}, \citenamefont {Neupert},
  \citenamefont {Kaminski}, \citenamefont {Lin}, \citenamefont {Jia},\ and\
  \citenamefont {{Zahid Hasan}}}]{Xu2015b}%
  \BibitemOpen
  \bibfield  {author} {\bibinfo {author} {\bibfnamefont {S.-Y.}\ \bibnamefont
  {Xu}}, \bibinfo {author} {\bibfnamefont {N.}~\bibnamefont {Alidoust}},
  \bibinfo {author} {\bibfnamefont {I.}~\bibnamefont {Belopolski}}, \bibinfo
  {author} {\bibfnamefont {Z.}~\bibnamefont {Yuan}}, \bibinfo {author}
  {\bibfnamefont {G.}~\bibnamefont {Bian}}, \bibinfo {author} {\bibfnamefont
  {T.-R.}\ \bibnamefont {Chang}}, \bibinfo {author} {\bibfnamefont
  {H.}~\bibnamefont {Zheng}}, \bibinfo {author} {\bibfnamefont {V.~N.}\
  \bibnamefont {Strocov}}, \bibinfo {author} {\bibfnamefont {D.~S.}\
  \bibnamefont {Sanchez}}, \bibinfo {author} {\bibfnamefont {G.}~\bibnamefont
  {Chang}}, \bibinfo {author} {\bibfnamefont {C.}~\bibnamefont {Zhang}},
  \bibinfo {author} {\bibfnamefont {D.}~\bibnamefont {Mou}}, \bibinfo {author}
  {\bibfnamefont {Y.}~\bibnamefont {Wu}}, \bibinfo {author} {\bibfnamefont
  {L.}~\bibnamefont {Huang}}, \bibinfo {author} {\bibfnamefont {C.-C.}\
  \bibnamefont {Lee}}, \bibinfo {author} {\bibfnamefont {S.-M.}\ \bibnamefont
  {Huang}}, \bibinfo {author} {\bibfnamefont {B.}~\bibnamefont {Wang}},
  \bibinfo {author} {\bibfnamefont {A.}~\bibnamefont {Bansil}}, \bibinfo
  {author} {\bibfnamefont {H.-T.}\ \bibnamefont {Jeng}}, \bibinfo {author}
  {\bibfnamefont {T.}~\bibnamefont {Neupert}}, \bibinfo {author} {\bibfnamefont
  {A.}~\bibnamefont {Kaminski}}, \bibinfo {author} {\bibfnamefont
  {H.}~\bibnamefont {Lin}}, \bibinfo {author} {\bibfnamefont {S.}~\bibnamefont
  {Jia}}, \ and\ \bibinfo {author} {\bibfnamefont {M.}~\bibnamefont {{Zahid
  Hasan}}},\ }\href {\doibase 10.1038/nphys3437} {\bibfield  {journal}
  {\bibinfo  {journal} {Nat. Phys.}\ }\textbf {\bibinfo {volume} {11}},\
  \bibinfo {pages} {748} (\bibinfo {year} {2015}{\natexlab{b}})}\BibitemShut
  {NoStop}%
\bibitem [{\citenamefont {Lv}\ \emph {et~al.}(2015)\citenamefont {Lv},
  \citenamefont {Weng}, \citenamefont {Fu}, \citenamefont {Wang}, \citenamefont
  {Miao}, \citenamefont {Ma}, \citenamefont {Richard}, \citenamefont {Huang},
  \citenamefont {Zhao}, \citenamefont {Chen}, \citenamefont {Fang},
  \citenamefont {Dai}, \citenamefont {Qian},\ and\ \citenamefont
  {Ding}}]{Lv2015}%
  \BibitemOpen
  \bibfield  {author} {\bibinfo {author} {\bibfnamefont {B.~Q.}\ \bibnamefont
  {Lv}}, \bibinfo {author} {\bibfnamefont {H.~M.}\ \bibnamefont {Weng}},
  \bibinfo {author} {\bibfnamefont {B.~B.}\ \bibnamefont {Fu}}, \bibinfo
  {author} {\bibfnamefont {X.~P.}\ \bibnamefont {Wang}}, \bibinfo {author}
  {\bibfnamefont {H.}~\bibnamefont {Miao}}, \bibinfo {author} {\bibfnamefont
  {J.}~\bibnamefont {Ma}}, \bibinfo {author} {\bibfnamefont {P.}~\bibnamefont
  {Richard}}, \bibinfo {author} {\bibfnamefont {X.~C.}\ \bibnamefont {Huang}},
  \bibinfo {author} {\bibfnamefont {L.~X.}\ \bibnamefont {Zhao}}, \bibinfo
  {author} {\bibfnamefont {G.~F.}\ \bibnamefont {Chen}}, \bibinfo {author}
  {\bibfnamefont {Z.}~\bibnamefont {Fang}}, \bibinfo {author} {\bibfnamefont
  {X.}~\bibnamefont {Dai}}, \bibinfo {author} {\bibfnamefont {T.}~\bibnamefont
  {Qian}}, \ and\ \bibinfo {author} {\bibfnamefont {H.}~\bibnamefont {Ding}},\
  }\href {\doibase 10.1103/PhysRevX.5.031013} {\bibfield  {journal} {\bibinfo
  {journal} {Phys. Rev. X}\ }\textbf {\bibinfo {volume} {5}},\ \bibinfo {pages}
  {031013} (\bibinfo {year} {2015})}\BibitemShut {NoStop}%
\bibitem [{\citenamefont {Borisenko}\ \emph {et~al.}(2014)\citenamefont
  {Borisenko}, \citenamefont {Gibson}, \citenamefont {Evtushinsky},
  \citenamefont {Zabolotnyy}, \citenamefont {B{\"{u}}chner},\ and\
  \citenamefont {Cava}}]{Borisenko2014}%
  \BibitemOpen
  \bibfield  {author} {\bibinfo {author} {\bibfnamefont {S.}~\bibnamefont
  {Borisenko}}, \bibinfo {author} {\bibfnamefont {Q.}~\bibnamefont {Gibson}},
  \bibinfo {author} {\bibfnamefont {D.}~\bibnamefont {Evtushinsky}}, \bibinfo
  {author} {\bibfnamefont {V.}~\bibnamefont {Zabolotnyy}}, \bibinfo {author}
  {\bibfnamefont {B.}~\bibnamefont {B{\"{u}}chner}}, \ and\ \bibinfo {author}
  {\bibfnamefont {R.~J.}\ \bibnamefont {Cava}},\ }\href {\doibase
  10.1103/PhysRevLett.113.027603} {\bibfield  {journal} {\bibinfo  {journal}
  {Phys. Rev. Lett.}\ }\textbf {\bibinfo {volume} {113}},\ \bibinfo {pages}
  {027603} (\bibinfo {year} {2014})}\BibitemShut {NoStop}%
\bibitem [{\citenamefont {Neupane}\ \emph {et~al.}(2014)\citenamefont
  {Neupane}, \citenamefont {Xu}, \citenamefont {Sankar}, \citenamefont
  {Alidoust}, \citenamefont {Bian}, \citenamefont {Liu}, \citenamefont
  {Belopolski}, \citenamefont {Chang}, \citenamefont {Jeng}, \citenamefont
  {Lin}, \citenamefont {Bansil}, \citenamefont {Chou},\ and\ \citenamefont
  {Hasan}}]{Neupane2014}%
  \BibitemOpen
  \bibfield  {author} {\bibinfo {author} {\bibfnamefont {M.}~\bibnamefont
  {Neupane}}, \bibinfo {author} {\bibfnamefont {S.-Y.}\ \bibnamefont {Xu}},
  \bibinfo {author} {\bibfnamefont {R.}~\bibnamefont {Sankar}}, \bibinfo
  {author} {\bibfnamefont {N.}~\bibnamefont {Alidoust}}, \bibinfo {author}
  {\bibfnamefont {G.}~\bibnamefont {Bian}}, \bibinfo {author} {\bibfnamefont
  {C.}~\bibnamefont {Liu}}, \bibinfo {author} {\bibfnamefont {I.}~\bibnamefont
  {Belopolski}}, \bibinfo {author} {\bibfnamefont {T.-R.}\ \bibnamefont
  {Chang}}, \bibinfo {author} {\bibfnamefont {H.-T.}\ \bibnamefont {Jeng}},
  \bibinfo {author} {\bibfnamefont {H.}~\bibnamefont {Lin}}, \bibinfo {author}
  {\bibfnamefont {A.}~\bibnamefont {Bansil}}, \bibinfo {author} {\bibfnamefont
  {F.}~\bibnamefont {Chou}}, \ and\ \bibinfo {author} {\bibfnamefont {M.~Z.}\
  \bibnamefont {Hasan}},\ }\href {\doibase 10.1038/ncomms4786} {\bibfield
  {journal} {\bibinfo  {journal} {Nat. Commun.}\ }\textbf {\bibinfo {volume}
  {5}},\ \bibinfo {pages} {3786} (\bibinfo {year} {2014})}\BibitemShut
  {NoStop}%
\bibitem [{\citenamefont {Liu}\ \emph {et~al.}(2014)\citenamefont {Liu},
  \citenamefont {Zhou}, \citenamefont {Zhang}, \citenamefont {Wang},
  \citenamefont {Weng}, \citenamefont {Prabhakaran}, \citenamefont {Mo},
  \citenamefont {Shen}, \citenamefont {Fang}, \citenamefont {Dai},
  \citenamefont {Hussain},\ and\ \citenamefont {Chen}}]{Liu2014a}%
  \BibitemOpen
  \bibfield  {author} {\bibinfo {author} {\bibfnamefont {Z.~K.}\ \bibnamefont
  {Liu}}, \bibinfo {author} {\bibfnamefont {B.}~\bibnamefont {Zhou}}, \bibinfo
  {author} {\bibfnamefont {Y.}~\bibnamefont {Zhang}}, \bibinfo {author}
  {\bibfnamefont {Z.~J.}\ \bibnamefont {Wang}}, \bibinfo {author}
  {\bibfnamefont {H.~M.}\ \bibnamefont {Weng}}, \bibinfo {author}
  {\bibfnamefont {D.}~\bibnamefont {Prabhakaran}}, \bibinfo {author}
  {\bibfnamefont {S.~K.}\ \bibnamefont {Mo}}, \bibinfo {author} {\bibfnamefont
  {Z.~X.}\ \bibnamefont {Shen}}, \bibinfo {author} {\bibfnamefont
  {Z.}~\bibnamefont {Fang}}, \bibinfo {author} {\bibfnamefont {X.}~\bibnamefont
  {Dai}}, \bibinfo {author} {\bibfnamefont {Z.}~\bibnamefont {Hussain}}, \ and\
  \bibinfo {author} {\bibfnamefont {Y.~L.}\ \bibnamefont {Chen}},\ }\href
  {\doibase 10.1126/science.1245085} {\bibfield  {journal} {\bibinfo  {journal}
  {Science}\ }\textbf {\bibinfo {volume} {343}},\ \bibinfo {pages} {864}
  (\bibinfo {year} {2014})}\BibitemShut {NoStop}%
\bibitem [{\citenamefont {Xiong}\ \emph {et~al.}(2015)\citenamefont {Xiong},
  \citenamefont {Kushwaha}, \citenamefont {Liang}, \citenamefont {Krizan},
  \citenamefont {Hirschberger}, \citenamefont {Wang}, \citenamefont {Cava},\
  and\ \citenamefont {Ong}}]{Xiong2015a}%
  \BibitemOpen
  \bibfield  {author} {\bibinfo {author} {\bibfnamefont {J.}~\bibnamefont
  {Xiong}}, \bibinfo {author} {\bibfnamefont {S.~K.}\ \bibnamefont {Kushwaha}},
  \bibinfo {author} {\bibfnamefont {T.}~\bibnamefont {Liang}}, \bibinfo
  {author} {\bibfnamefont {J.~W.}\ \bibnamefont {Krizan}}, \bibinfo {author}
  {\bibfnamefont {M.}~\bibnamefont {Hirschberger}}, \bibinfo {author}
  {\bibfnamefont {W.}~\bibnamefont {Wang}}, \bibinfo {author} {\bibfnamefont
  {R.~J.}\ \bibnamefont {Cava}}, \ and\ \bibinfo {author} {\bibfnamefont
  {N.~P.}\ \bibnamefont {Ong}},\ }\href {\doibase 10.1126/science.aac6089}
  {\bibfield  {journal} {\bibinfo  {journal} {Science}\ }\textbf {\bibinfo
  {volume} {350}},\ \bibinfo {pages} {413} (\bibinfo {year}
  {2015})}\BibitemShut {NoStop}%
\bibitem [{\citenamefont {Armitage}\ \emph {et~al.}(2018)\citenamefont
  {Armitage}, \citenamefont {Mele},\ and\ \citenamefont
  {Vishwanath}}]{Armitage2017}%
  \BibitemOpen
  \bibfield  {author} {\bibinfo {author} {\bibfnamefont {N.~P.}\ \bibnamefont
  {Armitage}}, \bibinfo {author} {\bibfnamefont {E.~J.}\ \bibnamefont {Mele}},
  \ and\ \bibinfo {author} {\bibfnamefont {A.}~\bibnamefont {Vishwanath}},\
  }\href {\doibase 10.1103/RevModPhys.90.015001} {\bibfield  {journal}
  {\bibinfo  {journal} {Rev. Mod. Phys.}\ }\textbf {\bibinfo {volume} {90}},\
  \bibinfo {pages} {015001} (\bibinfo {year} {2018})}\BibitemShut {NoStop}%
\bibitem [{\citenamefont {Yan}\ and\ \citenamefont {Felser}(2017)}]{Yan2017}%
  \BibitemOpen
  \bibfield  {author} {\bibinfo {author} {\bibfnamefont {B.}~\bibnamefont
  {Yan}}\ and\ \bibinfo {author} {\bibfnamefont {C.}~\bibnamefont {Felser}},\
  }\href {\doibase 10.1146/annurev-conmatphys-031016-025458} {\bibfield
  {journal} {\bibinfo  {journal} {Annu. Rev. Condens. Matter Phys.}\ }\textbf
  {\bibinfo {volume} {8}},\ \bibinfo {pages} {337} (\bibinfo {year}
  {2017})}\BibitemShut {NoStop}%
\bibitem [{\citenamefont {Nielsen}\ and\ \citenamefont
  {Ninomiya}(1983)}]{Nielsen1983}%
  \BibitemOpen
  \bibfield  {author} {\bibinfo {author} {\bibfnamefont {H.~B.}\ \bibnamefont
  {Nielsen}}\ and\ \bibinfo {author} {\bibfnamefont {M.}~\bibnamefont
  {Ninomiya}},\ }\href@noop {} {\bibfield  {journal} {\bibinfo  {journal}
  {Phys. Lett. B}\ }\textbf {\bibinfo {volume} {130(6)}},\ \bibinfo {pages}
  {389} (\bibinfo {year} {1983})}\BibitemShut {NoStop}%
\bibitem [{\citenamefont {Burkov}(2018)}]{Burkov2017}%
  \BibitemOpen
  \bibfield  {author} {\bibinfo {author} {\bibfnamefont {A.~A.}\ \bibnamefont
  {Burkov}},\ }\href {\doibase 10.1146/annurev-conmatphys-033117-054129}
  {\bibfield  {journal} {\bibinfo  {journal} {Annu. Rev. Condens. Matter
  Phys.}\ }\textbf {\bibinfo {volume} {9}},\ \bibinfo {pages} {359} (\bibinfo
  {year} {2018})}\BibitemShut {NoStop}%
\bibitem [{\citenamefont {Burkov}(2017)}]{Burkov2017a}%
  \BibitemOpen
  \bibfield  {author} {\bibinfo {author} {\bibfnamefont {A.~A.}\ \bibnamefont
  {Burkov}},\ }\href {\doibase 10.1103/PhysRevB.96.041110} {\bibfield
  {journal} {\bibinfo  {journal} {Phys. Rev. B}\ }\textbf {\bibinfo {volume}
  {96}},\ \bibinfo {pages} {041110(R)} (\bibinfo {year} {2017})}\BibitemShut
  {NoStop}%
\bibitem [{\citenamefont {Nandy}\ \emph {et~al.}(2017)\citenamefont {Nandy},
  \citenamefont {Sharma}, \citenamefont {Taraphder},\ and\ \citenamefont
  {Tewari}}]{Nandy2017}%
  \BibitemOpen
  \bibfield  {author} {\bibinfo {author} {\bibfnamefont {S.}~\bibnamefont
  {Nandy}}, \bibinfo {author} {\bibfnamefont {G.}~\bibnamefont {Sharma}},
  \bibinfo {author} {\bibfnamefont {A.}~\bibnamefont {Taraphder}}, \ and\
  \bibinfo {author} {\bibfnamefont {S.}~\bibnamefont {Tewari}},\ }\href
  {\doibase 10.1103/PhysRevLett.119.176804} {\bibfield  {journal} {\bibinfo
  {journal} {Phys. Rev. Lett}\ }\textbf {\bibinfo {volume} {119}},\ \bibinfo
  {pages} {176804} (\bibinfo {year} {2017})}\BibitemShut {NoStop}%
\bibitem [{\citenamefont {dos Reis}\ \emph {et~al.}(2016)\citenamefont {dos
  Reis}, \citenamefont {Ajeesh}, \citenamefont {Kumar}, \citenamefont {Arnold},
  \citenamefont {Shekhar}, \citenamefont {Naumann}, \citenamefont {Schmidt},
  \citenamefont {Nicklas},\ and\ \citenamefont {Hassinger}}]{Reis2016}%
  \BibitemOpen
  \bibfield  {author} {\bibinfo {author} {\bibfnamefont {R.~D.}\ \bibnamefont
  {dos Reis}}, \bibinfo {author} {\bibfnamefont {M.~O.}\ \bibnamefont
  {Ajeesh}}, \bibinfo {author} {\bibfnamefont {N.}~\bibnamefont {Kumar}},
  \bibinfo {author} {\bibfnamefont {F.}~\bibnamefont {Arnold}}, \bibinfo
  {author} {\bibfnamefont {C.}~\bibnamefont {Shekhar}}, \bibinfo {author}
  {\bibfnamefont {M.}~\bibnamefont {Naumann}}, \bibinfo {author} {\bibfnamefont
  {M.}~\bibnamefont {Schmidt}}, \bibinfo {author} {\bibfnamefont
  {M.}~\bibnamefont {Nicklas}}, \ and\ \bibinfo {author} {\bibfnamefont
  {E.}~\bibnamefont {Hassinger}},\ }\href {\doibase
  10.1088/1367-2630/18/8/085006} {\bibfield  {journal} {\bibinfo  {journal}
  {New J. Phys.}\ }\textbf {\bibinfo {volume} {18}},\ \bibinfo {pages} {085006}
  (\bibinfo {year} {2016})}\BibitemShut {NoStop}%
\bibitem [{\citenamefont {Qi}\ and\ \citenamefont {Zhang}(2011)}]{Qi2011}%
  \BibitemOpen
  \bibfield  {author} {\bibinfo {author} {\bibfnamefont {X.~L.}\ \bibnamefont
  {Qi}}\ and\ \bibinfo {author} {\bibfnamefont {S.~C.}\ \bibnamefont {Zhang}},\
  }\href {\doibase 10.1103/RevModPhys.83.1057} {\bibfield  {journal} {\bibinfo
  {journal} {Rev. Mod. Phys.}\ }\textbf {\bibinfo {volume} {83}},\ \bibinfo
  {pages} {1057} (\bibinfo {year} {2011})}\BibitemShut {NoStop}%
\bibitem [{\citenamefont {von Klitzing}(1986)}]{VonKlitzing1986}%
  \BibitemOpen
  \bibfield  {author} {\bibinfo {author} {\bibfnamefont {K.}~\bibnamefont {von
  Klitzing}},\ }\href {\doibase 10.1103/RevModPhys.58.519} {\bibfield
  {journal} {\bibinfo  {journal} {Rev. Mod. Phys.}\ }\textbf {\bibinfo {volume}
  {58}},\ \bibinfo {pages} {519} (\bibinfo {year} {1986})}\BibitemShut
  {NoStop}%
\bibitem [{\citenamefont {K{\"{o}}nig}\ \emph {et~al.}(2007)\citenamefont
  {K{\"{o}}nig}, \citenamefont {Wiedmann}, \citenamefont {Br{\"{u}}ne},
  \citenamefont {Roth}, \citenamefont {Buhmann}, \citenamefont {Molenkamp},
  \citenamefont {Qi},\ and\ \citenamefont {Zhang}}]{Konig2007}%
  \BibitemOpen
  \bibfield  {author} {\bibinfo {author} {\bibfnamefont {M.}~\bibnamefont
  {K{\"{o}}nig}}, \bibinfo {author} {\bibfnamefont {S.}~\bibnamefont
  {Wiedmann}}, \bibinfo {author} {\bibfnamefont {C.}~\bibnamefont
  {Br{\"{u}}ne}}, \bibinfo {author} {\bibfnamefont {A.}~\bibnamefont {Roth}},
  \bibinfo {author} {\bibfnamefont {H.}~\bibnamefont {Buhmann}}, \bibinfo
  {author} {\bibfnamefont {L.~W.}\ \bibnamefont {Molenkamp}}, \bibinfo {author}
  {\bibfnamefont {X.-L.}\ \bibnamefont {Qi}}, \ and\ \bibinfo {author}
  {\bibfnamefont {S.-C.}\ \bibnamefont {Zhang}},\ }\href {\doibase
  10.1126/science.1148047} {\bibfield  {journal} {\bibinfo  {journal}
  {Science}\ }\textbf {\bibinfo {volume} {318}},\ \bibinfo {pages} {766}
  (\bibinfo {year} {2007})}\BibitemShut {NoStop}%
\bibitem [{\citenamefont {Burkov}\ and\ \citenamefont
  {Balents}(2011)}]{Burkov2011}%
  \BibitemOpen
  \bibfield  {author} {\bibinfo {author} {\bibfnamefont {A.~A.}\ \bibnamefont
  {Burkov}}\ and\ \bibinfo {author} {\bibfnamefont {L.}~\bibnamefont
  {Balents}},\ }\href {\doibase https://doi.org/10.1103/PhysRevLett.107.127205}
  {\bibfield  {journal} {\bibinfo  {journal} {Phys. Rev. Lett}\ }\textbf
  {\bibinfo {volume} {107}},\ \bibinfo {pages} {127205} (\bibinfo {year}
  {2011})}\BibitemShut {NoStop}%
\bibitem [{\citenamefont {Burkov}(2014)}]{Burkov2014}%
  \BibitemOpen
  \bibfield  {author} {\bibinfo {author} {\bibfnamefont {A.~A.}\ \bibnamefont
  {Burkov}},\ }\href {\doibase 10.1103/PhysRevLett.113.187202} {\bibfield
  {journal} {\bibinfo  {journal} {Phys. Rev. Lett}\ }\textbf {\bibinfo {volume}
  {113}},\ \bibinfo {pages} {187202} (\bibinfo {year} {2014})}\BibitemShut
  {NoStop}%
\bibitem [{\citenamefont {Suzuki}\ \emph {et~al.}(2016)\citenamefont {Suzuki},
  \citenamefont {Chisnell}, \citenamefont {Devarakonda}, \citenamefont {Liu},
  \citenamefont {Feng}, \citenamefont {Xiao}, \citenamefont {Lynn},\ and\
  \citenamefont {Checkelsky}}]{Suzuki2016}%
  \BibitemOpen
  \bibfield  {author} {\bibinfo {author} {\bibfnamefont {T.}~\bibnamefont
  {Suzuki}}, \bibinfo {author} {\bibfnamefont {R.}~\bibnamefont {Chisnell}},
  \bibinfo {author} {\bibfnamefont {A.}~\bibnamefont {Devarakonda}}, \bibinfo
  {author} {\bibfnamefont {Y.-T.}\ \bibnamefont {Liu}}, \bibinfo {author}
  {\bibfnamefont {W.}~\bibnamefont {Feng}}, \bibinfo {author} {\bibfnamefont
  {D.}~\bibnamefont {Xiao}}, \bibinfo {author} {\bibfnamefont {J.~W.}\
  \bibnamefont {Lynn}}, \ and\ \bibinfo {author} {\bibfnamefont {J.~G.}\
  \bibnamefont {Checkelsky}},\ }\href {\doibase 10.1038/nphys3831} {\bibfield
  {journal} {\bibinfo  {journal} {Nat. Phys.}\ }\textbf {\bibinfo {volume}
  {12}},\ \bibinfo {pages} {1119} (\bibinfo {year} {2016})}\BibitemShut
  {NoStop}%
\bibitem [{\citenamefont {Liu}\ \emph {et~al.}(2018)\citenamefont {Liu},
  \citenamefont {Sun}, \citenamefont {M{\"{u}}chler}, \citenamefont {Sun},
  \citenamefont {Jiao}, \citenamefont {Kroder}, \citenamefont {S{\"{u}}{\ss}},
  \citenamefont {Borrmann}, \citenamefont {Wang}, \citenamefont {Schnelle},
  \citenamefont {Wirth}, \citenamefont {Goennenwein},\ and\ \citenamefont
  {Felser}}]{Liu2017c}%
  \BibitemOpen
  \bibfield  {author} {\bibinfo {author} {\bibfnamefont {E.}~\bibnamefont
  {Liu}}, \bibinfo {author} {\bibfnamefont {Y.}~\bibnamefont {Sun}}, \bibinfo
  {author} {\bibfnamefont {L.}~\bibnamefont {M{\"{u}}chler}}, \bibinfo {author}
  {\bibfnamefont {A.}~\bibnamefont {Sun}}, \bibinfo {author} {\bibfnamefont
  {L.}~\bibnamefont {Jiao}}, \bibinfo {author} {\bibfnamefont {J.}~\bibnamefont
  {Kroder}}, \bibinfo {author} {\bibfnamefont {V.}~\bibnamefont
  {S{\"{u}}{\ss}}}, \bibinfo {author} {\bibfnamefont {H.}~\bibnamefont
  {Borrmann}}, \bibinfo {author} {\bibfnamefont {W.}~\bibnamefont {Wang}},
  \bibinfo {author} {\bibfnamefont {W.}~\bibnamefont {Schnelle}}, \bibinfo
  {author} {\bibfnamefont {S.}~\bibnamefont {Wirth}}, \bibinfo {author}
  {\bibfnamefont {S.~T.~B.}\ \bibnamefont {Goennenwein}}, \ and\ \bibinfo
  {author} {\bibfnamefont {C.}~\bibnamefont {Felser}},\ }\href {\doibase
  https://doi.org/10.1038/s41567-018-0234-5} {\bibfield  {journal} {\bibinfo
  {journal} {Nat. Phys.}\ }\textbf {\bibinfo {volume} {14}},\ \bibinfo {pages}
  {1125} (\bibinfo {year} {2018})}\BibitemShut {NoStop}%
\bibitem [{\citenamefont {Li}\ \emph {et~al.}()\citenamefont {Li},
  \citenamefont {Koo}, \citenamefont {Ning}, \citenamefont {Li}, \citenamefont
  {Miao}, \citenamefont {Min}, \citenamefont {Zhu}, \citenamefont {Wang},
  \citenamefont {Alem}, \citenamefont {Liu}, \citenamefont {Mao},\ and\
  \citenamefont {Yan}}]{Li2019}%
  \BibitemOpen
  \bibfield  {author} {\bibinfo {author} {\bibfnamefont {P.}~\bibnamefont
  {Li}}, \bibinfo {author} {\bibfnamefont {J.}~\bibnamefont {Koo}}, \bibinfo
  {author} {\bibfnamefont {W.}~\bibnamefont {Ning}}, \bibinfo {author}
  {\bibfnamefont {J.}~\bibnamefont {Li}}, \bibinfo {author} {\bibfnamefont
  {L.}~\bibnamefont {Miao}}, \bibinfo {author} {\bibfnamefont {L.}~\bibnamefont
  {Min}}, \bibinfo {author} {\bibfnamefont {Y.}~\bibnamefont {Zhu}}, \bibinfo
  {author} {\bibfnamefont {Y.}~\bibnamefont {Wang}}, \bibinfo {author}
  {\bibfnamefont {N.}~\bibnamefont {Alem}}, \bibinfo {author} {\bibfnamefont
  {C.-X.}\ \bibnamefont {Liu}}, \bibinfo {author} {\bibfnamefont
  {Z.}~\bibnamefont {Mao}}, \ and\ \bibinfo {author} {\bibfnamefont
  {B.}~\bibnamefont {Yan}},\ }\href@noop {} {\ }\Eprint
  {http://arxiv.org/abs/1910.10378} {arXiv:1910.10378} \BibitemShut {NoStop}%
\bibitem [{\citenamefont {Balents}(2011)}]{Balents2011}%
  \BibitemOpen
  \bibfield  {author} {\bibinfo {author} {\bibfnamefont {L.}~\bibnamefont
  {Balents}},\ }\href {\doibase 10.1103/Physics.4.36} {\bibfield  {journal}
  {\bibinfo  {journal} {Physics (College. Park. Md).}\ }\textbf {\bibinfo
  {volume} {4}},\ \bibinfo {pages} {36} (\bibinfo {year} {2011})}\BibitemShut
  {NoStop}%
\bibitem [{\citenamefont {Breitkreiz}\ and\ \citenamefont
  {Brouwer}(2019)}]{Breitkreiz2019}%
  \BibitemOpen
  \bibfield  {author} {\bibinfo {author} {\bibfnamefont {M.}~\bibnamefont
  {Breitkreiz}}\ and\ \bibinfo {author} {\bibfnamefont {P.~W.}\ \bibnamefont
  {Brouwer}},\ }\href {\doibase 10.1103/PhysRevLett.123.066804} {\bibfield
  {journal} {\bibinfo  {journal} {Phys. Rev. Lett}\ }\textbf {\bibinfo {volume}
  {123}},\ \bibinfo {pages} {066804} (\bibinfo {year} {2019})}\BibitemShut
  {NoStop}%
\bibitem [{\citenamefont {Colom{\'{e}}s}\ and\ \citenamefont
  {Franz}(2018)}]{Colomes2018}%
  \BibitemOpen
  \bibfield  {author} {\bibinfo {author} {\bibfnamefont {E.}~\bibnamefont
  {Colom{\'{e}}s}}\ and\ \bibinfo {author} {\bibfnamefont {M.}~\bibnamefont
  {Franz}},\ }\href {\doibase 10.1103/PhysRevLett.120.086603} {\bibfield
  {journal} {\bibinfo  {journal} {Phys. Rev. Lett.}\ }\textbf {\bibinfo
  {volume} {120}},\ \bibinfo {pages} {086603} (\bibinfo {year}
  {2018})}\BibitemShut {NoStop}%
\bibitem [{\citenamefont {Mandal}\ \emph {et~al.}(2019)\citenamefont {Mandal},
  \citenamefont {Ge},\ and\ \citenamefont {Liew}}]{Mandal2019}%
  \BibitemOpen
  \bibfield  {author} {\bibinfo {author} {\bibfnamefont {S.}~\bibnamefont
  {Mandal}}, \bibinfo {author} {\bibfnamefont {R.}~\bibnamefont {Ge}}, \ and\
  \bibinfo {author} {\bibfnamefont {T.~C.}\ \bibnamefont {Liew}},\ }\href
  {\doibase 10.1103/PhysRevB.99.115423} {\bibfield  {journal} {\bibinfo
  {journal} {Phys. Rev. B}\ }\textbf {\bibinfo {volume} {99}},\ \bibinfo
  {pages} {115423} (\bibinfo {year} {2019})}\BibitemShut {NoStop}%
\bibitem [{\citenamefont {Baireuther}\ \emph {et~al.}(2016)\citenamefont
  {Baireuther}, \citenamefont {Hutasoit}, \citenamefont {Tworzyd{\l}o},\ and\
  \citenamefont {Beenakker}}]{Baireuther2016}%
  \BibitemOpen
  \bibfield  {author} {\bibinfo {author} {\bibfnamefont {P.}~\bibnamefont
  {Baireuther}}, \bibinfo {author} {\bibfnamefont {J.~A.}\ \bibnamefont
  {Hutasoit}}, \bibinfo {author} {\bibfnamefont {J.}~\bibnamefont
  {Tworzyd{\l}o}}, \ and\ \bibinfo {author} {\bibfnamefont {C.~W.~J.}\
  \bibnamefont {Beenakker}},\ }\href {\doibase 0.1088/1367-2630/18/4/045009}
  {\bibfield  {journal} {\bibinfo  {journal} {New J. Phys}\ }\textbf {\bibinfo
  {volume} {18}},\ \bibinfo {pages} {045009} (\bibinfo {year}
  {2016})}\BibitemShut {NoStop}%
\bibitem [{\citenamefont {Pikulin}\ \emph {et~al.}(2016)\citenamefont
  {Pikulin}, \citenamefont {Chen},\ and\ \citenamefont {Franz}}]{Pikulin2016}%
  \BibitemOpen
  \bibfield  {author} {\bibinfo {author} {\bibfnamefont {D.~I.}\ \bibnamefont
  {Pikulin}}, \bibinfo {author} {\bibfnamefont {A.}~\bibnamefont {Chen}}, \
  and\ \bibinfo {author} {\bibfnamefont {M.}~\bibnamefont {Franz}},\ }\href
  {\doibase 10.1103/PhysRevX.6.041021} {\bibfield  {journal} {\bibinfo
  {journal} {Phys. Rev. X}\ }\textbf {\bibinfo {volume} {6}},\ \bibinfo {pages}
  {041021} (\bibinfo {year} {2016})}\BibitemShut {NoStop}%
\bibitem [{\citenamefont {Sodemann}\ and\ \citenamefont
  {Fu}(2015)}]{Sodemann2015}%
  \BibitemOpen
  \bibfield  {author} {\bibinfo {author} {\bibfnamefont {I.}~\bibnamefont
  {Sodemann}}\ and\ \bibinfo {author} {\bibfnamefont {L.}~\bibnamefont {Fu}},\
  }\href {\doibase 10.1103/PhysRevLett.115.216806} {\bibfield  {journal}
  {\bibinfo  {journal} {Phys. Rev. Lett.}\ }\textbf {\bibinfo {volume} {115}},\
  \bibinfo {pages} {216806} (\bibinfo {year} {2015})}\BibitemShut {NoStop}%
\bibitem [{\citenamefont {Bovenzi}\ \emph {et~al.}(2018)\citenamefont
  {Bovenzi}, \citenamefont {Breitkreiz}, \citenamefont {O'Brien}, \citenamefont
  {Tworzyd{\l}o},\ and\ \citenamefont {Beenakker}}]{Bovenzi2017a}%
  \BibitemOpen
  \bibfield  {author} {\bibinfo {author} {\bibfnamefont {N.}~\bibnamefont
  {Bovenzi}}, \bibinfo {author} {\bibfnamefont {M.}~\bibnamefont {Breitkreiz}},
  \bibinfo {author} {\bibfnamefont {T.~E.}\ \bibnamefont {O'Brien}}, \bibinfo
  {author} {\bibfnamefont {J.}~\bibnamefont {Tworzyd{\l}o}}, \ and\ \bibinfo
  {author} {\bibfnamefont {C.~W.~J.}\ \bibnamefont {Beenakker}},\ }\href
  {\doibase 10.1088/1367-2630/aaaa90} {\bibfield  {journal} {\bibinfo
  {journal} {New J. Phys.}\ }\textbf {\bibinfo {volume} {20}},\ \bibinfo
  {pages} {023023} (\bibinfo {year} {2018})}\BibitemShut {NoStop}%
\bibitem [{\citenamefont {Mahan}(2000)}]{Mahan2000}%
  \BibitemOpen
  \bibfield  {author} {\bibinfo {author} {\bibfnamefont {G.~D.}\ \bibnamefont
  {Mahan}},\ }\href@noop {} {\emph {\bibinfo {title} {{Many-Particle
  Physics}}}}\ (\bibinfo  {publisher} {Kluwer Academic},\ \bibinfo {address}
  {New York},\ \bibinfo {year} {2000})\BibitemShut {NoStop}%
\bibitem [{dum({\natexlab{a}})}]{dummy2}%
  \BibitemOpen
  \href@noop {} {\ }\BibitemShut {NoStop}%
\bibitem [{\citenamefont {Kleiner}(1966)}]{Kleiner1966}%
  \BibitemOpen
  \bibfield  {author} {\bibinfo {author} {\bibfnamefont {W.~H.}\ \bibnamefont
  {Kleiner}},\ }\href {\doibase 10.1103/PhysRev.142.318} {\bibfield  {journal}
  {\bibinfo  {journal} {Phys. Rev.}\ }\textbf {\bibinfo {volume} {142}},\
  \bibinfo {pages} {318} (\bibinfo {year} {1966})}\BibitemShut {NoStop}%
\bibitem [{\citenamefont {Seemann}\ \emph {et~al.}(2015)\citenamefont
  {Seemann}, \citenamefont {K{\"{o}}dderitzsch}, \citenamefont {Wimmer},\ and\
  \citenamefont {Ebert}}]{Seemann2015}%
  \BibitemOpen
  \bibfield  {author} {\bibinfo {author} {\bibfnamefont {M.}~\bibnamefont
  {Seemann}}, \bibinfo {author} {\bibfnamefont {D.}~\bibnamefont
  {K{\"{o}}dderitzsch}}, \bibinfo {author} {\bibfnamefont {S.}~\bibnamefont
  {Wimmer}}, \ and\ \bibinfo {author} {\bibfnamefont {H.}~\bibnamefont
  {Ebert}},\ }\href {\doibase 10.1103/PhysRevB.92.155138} {\bibfield  {journal}
  {\bibinfo  {journal} {Phys. Rev. B}\ }\textbf {\bibinfo {volume} {92}},\
  \bibinfo {pages} {155138} (\bibinfo {year} {2015})}\BibitemShut {NoStop}%
\bibitem [{\citenamefont {Parameswaran}\ \emph {et~al.}(2014)\citenamefont
  {Parameswaran}, \citenamefont {Grover}, \citenamefont {Abanin}, \citenamefont
  {Pesin},\ and\ \citenamefont {Vishwanath}}]{Parameswaran2014}%
  \BibitemOpen
  \bibfield  {author} {\bibinfo {author} {\bibfnamefont {S.~A.}\ \bibnamefont
  {Parameswaran}}, \bibinfo {author} {\bibfnamefont {T.}~\bibnamefont
  {Grover}}, \bibinfo {author} {\bibfnamefont {D.~A.}\ \bibnamefont {Abanin}},
  \bibinfo {author} {\bibfnamefont {D.~A.}\ \bibnamefont {Pesin}}, \ and\
  \bibinfo {author} {\bibfnamefont {A.}~\bibnamefont {Vishwanath}},\ }\href
  {\doibase 10.1103/PhysRevX.4.031035} {\bibfield  {journal} {\bibinfo
  {journal} {Phys. Rev. X}\ }\textbf {\bibinfo {volume} {4}},\ \bibinfo {pages}
  {031035} (\bibinfo {year} {2014})}\BibitemShut {NoStop}%
\bibitem [{\citenamefont {Behrends}\ \emph {et~al.}(2019)\citenamefont
  {Behrends}, \citenamefont {Ilan},\ and\ \citenamefont
  {Bardarson}}]{Behrends2019}%
  \BibitemOpen
  \bibfield  {author} {\bibinfo {author} {\bibfnamefont {J.}~\bibnamefont
  {Behrends}}, \bibinfo {author} {\bibfnamefont {R.}~\bibnamefont {Ilan}}, \
  and\ \bibinfo {author} {\bibfnamefont {J.~H.}\ \bibnamefont {Bardarson}},\
  }\href {\doibase 10.1103/PhysRevResearch.1.032028} {\bibfield  {journal}
  {\bibinfo  {journal} {Phys. Rev. Res.}\ }\textbf {\bibinfo {volume} {1}},\
  \bibinfo {pages} {032028} (\bibinfo {year} {2019})}\BibitemShut {NoStop}%
\bibitem [{dum({\natexlab{b}})}]{dummy}%
  \BibitemOpen
  \href@noop {} {\  }\BibitemShut {NoStop}%
\bibitem [{\citenamefont {Grushin}\ \emph {et~al.}(2016)\citenamefont
  {Grushin}, \citenamefont {Venderbos}, \citenamefont {Vishwanath},\ and\
  \citenamefont {Ilan}}]{Grushin2016}%
  \BibitemOpen
  \bibfield  {author} {\bibinfo {author} {\bibfnamefont {A.~G.}\ \bibnamefont
  {Grushin}}, \bibinfo {author} {\bibfnamefont {J.~W.~F.}\ \bibnamefont
  {Venderbos}}, \bibinfo {author} {\bibfnamefont {A.}~\bibnamefont
  {Vishwanath}}, \ and\ \bibinfo {author} {\bibfnamefont {R.}~\bibnamefont
  {Ilan}},\ }\href {\doibase 10.1103/PhysRevX.6.041046} {\bibfield  {journal}
  {\bibinfo  {journal} {Phys. Rev. X}\ }\textbf {\bibinfo {volume} {6}},\
  \bibinfo {pages} {041046} (\bibinfo {year} {2016})}\BibitemShut {NoStop}%
\bibitem [{\citenamefont {Pereira}\ \emph {et~al.}(2019)\citenamefont
  {Pereira}, \citenamefont {Buccheri}, \citenamefont {{De Martino}},\ and\
  \citenamefont {Egger}}]{Pereira2019}%
  \BibitemOpen
  \bibfield  {author} {\bibinfo {author} {\bibfnamefont {R.~G.}\ \bibnamefont
  {Pereira}}, \bibinfo {author} {\bibfnamefont {F.}~\bibnamefont {Buccheri}},
  \bibinfo {author} {\bibfnamefont {A.}~\bibnamefont {{De Martino}}}, \ and\
  \bibinfo {author} {\bibfnamefont {R.}~\bibnamefont {Egger}},\ }\href
  {\doibase 10.1103/PhysRevB.100.035106} {\bibfield  {journal} {\bibinfo
  {journal} {Phys. Rev. B}\ }\textbf {\bibinfo {volume} {100}},\ \bibinfo
  {pages} {035106} (\bibinfo {year} {2019})}\BibitemShut {NoStop}%
\bibitem [{\citenamefont {Zhang}\ \emph {et~al.}(2019)\citenamefont {Zhang},
  \citenamefont {Ni}, \citenamefont {Zhang}, \citenamefont {Yuan},
  \citenamefont {Liu}, \citenamefont {Zou}, \citenamefont {Liao}, \citenamefont
  {Du}, \citenamefont {Narayan}, \citenamefont {Zhang}, \citenamefont {Gu},
  \citenamefont {Zhu}, \citenamefont {Pi}, \citenamefont {Sanvito},
  \citenamefont {Han}, \citenamefont {Zou}, \citenamefont {Shi}, \citenamefont
  {Wan}, \citenamefont {Savrasov},\ and\ \citenamefont {Xiu}}]{Zhang2019}%
  \BibitemOpen
  \bibfield  {author} {\bibinfo {author} {\bibfnamefont {C.}~\bibnamefont
  {Zhang}}, \bibinfo {author} {\bibfnamefont {Z.}~\bibnamefont {Ni}}, \bibinfo
  {author} {\bibfnamefont {J.}~\bibnamefont {Zhang}}, \bibinfo {author}
  {\bibfnamefont {X.}~\bibnamefont {Yuan}}, \bibinfo {author} {\bibfnamefont
  {Y.}~\bibnamefont {Liu}}, \bibinfo {author} {\bibfnamefont {Y.}~\bibnamefont
  {Zou}}, \bibinfo {author} {\bibfnamefont {Z.}~\bibnamefont {Liao}}, \bibinfo
  {author} {\bibfnamefont {Y.}~\bibnamefont {Du}}, \bibinfo {author}
  {\bibfnamefont {A.}~\bibnamefont {Narayan}}, \bibinfo {author} {\bibfnamefont
  {H.}~\bibnamefont {Zhang}}, \bibinfo {author} {\bibfnamefont
  {T.}~\bibnamefont {Gu}}, \bibinfo {author} {\bibfnamefont {X.}~\bibnamefont
  {Zhu}}, \bibinfo {author} {\bibfnamefont {L.}~\bibnamefont {Pi}}, \bibinfo
  {author} {\bibfnamefont {S.}~\bibnamefont {Sanvito}}, \bibinfo {author}
  {\bibfnamefont {X.}~\bibnamefont {Han}}, \bibinfo {author} {\bibfnamefont
  {J.}~\bibnamefont {Zou}}, \bibinfo {author} {\bibfnamefont {Y.}~\bibnamefont
  {Shi}}, \bibinfo {author} {\bibfnamefont {X.}~\bibnamefont {Wan}}, \bibinfo
  {author} {\bibfnamefont {S.~Y.}\ \bibnamefont {Savrasov}}, \ and\ \bibinfo
  {author} {\bibfnamefont {F.}~\bibnamefont {Xiu}},\ }\href {\doibase
  10.1038/s41563-019-0320-9} {\bibfield  {journal} {\bibinfo  {journal} {Nat.
  Mater.}\ }\textbf {\bibinfo {volume} {18}},\ \bibinfo {pages} {482} (\bibinfo
  {year} {2019})}\BibitemShut {NoStop}%
\end{thebibliography}%

\clearpage

\onecolumngrid

\renewcommand{\theequation}{S\arabic{equation}}
\renewcommand{\thefigure}{S\arabic{figure}}

\setcounter{equation}{0}
\setcounter{figure}{0}

\section*{Supplemental Material}

\subsection{Quantum Boltzmann formalism}

To explore the transport behavior in linear response, but 
at the same time allowing non-linear dependencies on the
spatial coordinate, we use the 
Quantum Boltzmann Transport formalism  \cite{Mahan2000}
allowing an arbitrary spatial variation 
 of the Wigner distribution function at state $\vk$,
$f(\vk,\vx,\omega)$, where 
$\vx$ denotes the three-dimensional center-of-mass position. 
In this inhomogeneous case, an 
external field, inducing the steady non-equilibrium
state, can be incorporated via
non-equilibrium boundary conditions on the 
spatial variation of the distribution, hence no 
explicit external field will be needed. Assuming impurity 
scattering in first Born approximation
($T$ matrix given by impurity-potential matrix elements
$V_{\vk\vk'}$) and assuming that the associated
renormalization of the dispersion relation is already
incorporated in $\varepsilon_\vk$, the kinetic equation reads 
\begin{align}
 &\vv_\vk\cdot\vn f(\vk,\vx,\omega)
= A(\vk,\vx,\omega)\int \frac{d^3\kappa'}{(2\pi)^3} |V_{\vk\vk'}|^2
f(\vk', \vx,\omega)
- f(\vk, \vx,\omega)\int \frac{d^3 \kappa'}{(2\pi)^3} |V_{\vk\vk'}|^2 A(\vk', \vx,\omega).\label{ke}
\end{align}
Following the standard route, from the Wigner distribution function
$f(\vk, \vx,\omega)$ we 
factor out the spectral density $A(\vk, \vx,\omega)$ such that
$f(\vk, \vx,\omega)= f(\vk, \vx)A(\vk,\vx,\omega)$, defining
the non-equilibrium occupation function $f(\vk,\vx)$. 
Assuming weak scattering, we employ the quasiparticle 
approximation and use the free-particle expression  
for the spectral function (from the main text),
\begin{align}
A(\vk, \vx,\omega) &=  \delta(\omega-\varepsilon_\vk)\rho(\vk, x),
 \label{A}\\
\rho(\vk, x) &= 
 \begin{cases} 
1 &  \mathrm{Im}\, q = 0 \\   
2Wm(k_z)\; e^{\pm 2m(k_z) (x\mp W/2)} & \mathrm{Im}\, q \neq 0
\end{cases} 
\label{rho}
\end{align}
  Inserting
 into \eqref{ke} and integrating over $\omega$, we obtain
 the Boltzmann equation 
\begin{equation}
\vv_\vk\cdot\vn f(\vk, \vx) 
=\sum_{\vk'} \Gamma_{\vk\vk'}\big[ f(\vk', \vx) - f(\vk, \vx)\big], \label{be2}
\end{equation}
where we divided both sides by $\rho(\vk, x)$ 
(keeping in mind that  Eq.\ \eqref{be2} becomes
trivial if $\rho(\vk, x)= 0$, i.e., for surface states 
away from the boundary). In the Boltzmann Equation
 we defined
\begin{equation}
\sum_{\vk'} \Gamma_{\vk\vk'}\dots= \int \frac{d^3 \kappa'}{(2\pi)^3} |V_{\vk\vk'}|^2
\delta(\varepsilon_\vk-\varepsilon_{\vk'})\rho(\vk',x)\dots.
\end{equation}
To focus more deeply on 
qualitative features, we assume that the 
scattering amplitudes are 
different only between states of the different types, so that 
$\Gamma_{\vk\vk'} \to \Gamma_{ij}$, where $i,j \in[ n+,c+,s+,n-,c-,s-]$. 
In this case, the integral over states naturally splits into 
integrals over the different particle types at the Fermi level,
\begin{align}
\sum_{\vk'} =& \sum_{\vk'}^{n\pm} 
+\sum_{\vk'}^{c\pm} +s(x)\sum_{\vk'}^{s},\nonumber\\
&\sum_{\vk'}^{n\pm} = \int^{n\pm} \frac{d^2\kappa'}{4\pi^2 h v},
\;\;\;\;\;
\sum_{\vk'}^{c\pm} = \frac{1}{W}\int^{c\pm}  \frac{d \kappa '}{2\pi h v},\;\;\;\;\;
\sum_{\vk'}^{s} = \int^{s\pm}  \frac{d \kappa'}{2\pi h v} ,
\end{align}
where $s(x) = \sum_\pm \delta(x\pm W/2) $. 
Since the Boltzmann equation  \eqref{be2} 
is non-trivial only for the non-equilibrium part 
of the occupation, 
the usual ansatz $f(\vk, \vx) = n_F(\varepsilon_\vk)+
n_F'(\varepsilon_\vk) \mu(\vk, \vx)$, 
where $ n_F(\varepsilon_\vk)$ is the equilibrium 
occupation function,  replaces $f(\vk, \vx)\to\mu(\vk, \vx)$
in \eqref{be2}. We define the 
non-equilibrium chemical potential
and the particle-current density  of each particle type
$i$, respectively, as 
\begin{equation}
\mu_{i} = \langle \mu(\vk)\rangle_i =
 \sum_{\vk}^{i} \mu(\vk, \vx) \big/  n_i,\;\;\;\;\;
\vj_{i\pm} =n_i\langle {\bm v}_\vk\mu(\vk, \vx)\rangle_i =\sum_{\vk}^{i} \vv_\vk  \mu(\vk, \vx), \label{cds}
\end{equation}
where $n_i = \sum_{\vk}^{i}$ is the density of states of the 
particle type $i$. Defining the diffusion constant of bulk 
particles as $
D = v^2\big/ 3 \sum_{\vk'}\Gamma_{n\pm\, \vk'}$
and summing the Boltzmann equation  \eqref{be2} over  
$n\pm$ after multiplying with $\vv_\vk$ and assuming the Weyl
Fermi surfaces to be spherical, we obtain
\begin{align}
\vj_{n\pm} &= -n_{\pm}D\, \vn\mu_{n\pm}.
 \label{jnpms}
\end{align}
Summing the Boltzmann equation  \eqref{be2} over each particle type gives
\begin{subequations}
\begin{align}
\vn\cdot\vj_{n\pm} = & \pm (\gamma_{c\mp n\pm}
+ \gamma_{n- n+})(\mu_{n-}-\mu_{n+})
 +\gamma_{sn\pm}s(x) (\mu_s-\mu_{n\pm}), \\
\vn \cdot\vj_{c\pm} =& \pm \gamma_{n\mp c\pm}
 (\mu_{n-}-\mu_{n+}) 
+ \gamma_{sc\pm}s(x) (\mu_s-\mu_{c\pm}), \\
s(x) \vn\cdot\vj_{s\pm} =&  s(x)\big[
\gamma_{sn+}(\mu_{n+}-\mu_{s\pm}) 
+\gamma_{sn-}(\mu_{n-}-\mu_{s\pm})
+\gamma_{sc+}(\mu_{c+}-\mu_{s\pm}) 
+\gamma_{sc-}(\mu_{c-}-\mu_{s\pm}) \big],
\end{align}
\label{eqss}
\end{subequations}
where $\gamma_{ij} =\Gamma_{ij}n_in_j$ and 
we assumed that intracone scattering is much larger than
intercone and bulk-surface scattering so that
$\mu_{n\pm}-\mu_{c\pm} \approx 0$ away from the 
boundary. Taking into account that the number of normal 
bulk states $n\pm$ strongly dominates since $W\gg 1$
the equations simplify to Eqs.\ (11) in the main text.

\subsection{Solving the transport equations}

We now solve for the case of a homogeneous 
force field applied in the
$z$ direction in form of a potential gradient $\partial_z \mu = E$.
To linear order in $E$ and using translation invariance
in the $y$ direction, the divergence of the 
chiral and the surface particle currents simplify to
\begin{align}
\vn\cdot\vj_{c\pm} &= \bar{nv}_{c\pm} E ,
\;\;\;\;\;\; 
\vn\cdot \vj_{s\pm} =\bar{nv}_s^z E, \label{jcpms} 
\end{align}
where 
the density-of-states
weighted integral over the velocity
at the Fermi level reads
\begin{equation}
\bar{nv}_{c\pm} = \pm\frac{\varepsilon_F\mp \varepsilon_c}{\pi h W},
\;\;\;\;\;\bar{nv}^z_{s} = \frac{\varepsilon_c}{\pi h} = \frac{\varepsilon_c}{k_0} n_s v. \label{nvs}
\end{equation}
A solution for the non-equilibrium potential profile in the
$x$ direction is 
found by first solving for the non-equilibrium potential for  
the bulk. Away from the boundary ($s(x)=0$) we obtian 
from \eqref{eqss},  \eqref{jnpms},  and \eqref{jcpms} 
\begin{equation}
\bar{nv}_{c\pm} E
-n_{n\pm}D \partial^2_x \mu_{n\pm} = \pm
 \frac{n_+n_-}{n_n} \frac{D}{l^2_c} 
(\mu_{n-}-\mu_{n+}), \label{eqbs}
\end{equation}
where we used $\gamma_{c-n+}+\gamma_{n-n+}+\gamma_{n-c+}\approx n_+n_- \Gamma_{n+n-}$ since $n_{c\pm}\ll 
n_\pm$ and defined the 
chiral relaxation length $l_c \equiv \sqrt{D/\Gamma_{n+n-} n_n}$.
Boundary conditions for $\mu_{n\pm}$ are given by integration 
of \eqref{eqss} over an infinitesimal distance at both boundaries and assuming vanishing current through the boundary,
\begin{subequations}
\begin{align}
j^x_{n+}(\pm W/2) =& \mp
\gamma_{sn+}(\mu_{s\pm}-\mu_{n+}(\pm W/2)), \\
j^x_{n-}(\pm W/2) =& \mp
\gamma_{sn-} (\mu_{s\pm}-\mu_{n-}(\pm W/2)), \\
E\bar{nv}^z_s=&   
\gamma_{sn+} (\mu_{n+}(\pm W/2)-\mu_{s\pm}) +\gamma_{sn-} (\mu_{n-}(\pm W/2)-\mu_{s\pm}).
\label{muss}
\end{align}
\label{bcss}
\end{subequations}
We now introduce the bulk chemical  potential $\mu_n$,
which is the local chemical potential averaged over
all bulk states,
and the chiral chemical potential $\mu_a$, 
\begin{equation}
\mu_{n/a} = \frac{n_{n+}\mu_{n+}\pm n_{n-}\mu_{n-}}{n_n} \;\;\;\;\; \Leftrightarrow\;\;\;\;\;
\mu_{n\pm} = n_n\frac{\mu_n\pm \mu_a}{2n_\pm}.
\label{muna}
\end{equation}
In terms of $\mu_{n}$ and $\mu_{a}$,  
Eqs. \eqref{eqbs} and the boundary 
conditions \eqref{bcss} partially decouple into 
\begin{subequations}
\begin{align}
\partial_x^2\mu_n &= -2 \theta \frac{\varepsilon_c}{k_0}
\,\frac{E}{W}\\
\partial\mu_n\big|_{x=\pm W/2} & = \mp \theta \frac{\varepsilon_c}{k_0}\, \frac{E}{W}\\
\partial_x^2\mu_a &=\frac{1}{l^2_c}\big[\mu_a 
-\gamma \mu_n\big]+2\theta\frac{\varepsilon_F}{k_0}\frac{E}{W} , \\
\partial\mu_a\big|_{x=\pm W/2} &= \mp\bigg[
\theta\frac{\varepsilon_c}{k_0}\gamma\Big( 1+ \frac{\theta W}{4l_{ns}}\Big) E +
\frac{\theta}{l_{ns}}\mu_a\big|_{x=\pm W/2}\bigg],
\end{align}
\label{eq2s}
\end{subequations}
where $\gamma=(n_+-n_-)/n_n$, 
  $l_{ns}= v/\Gamma_{sn}n_n$ is the scattering length of
surface states, and
\begin{equation}
\theta=\frac{n_s v}{n_nD} = \frac{v}{2D} \frac{k_0}{
\varepsilon_c^2+\varepsilon_F^2}
\end{equation}
 is the Hall angle of the AHE associated with the system of two Weyl nodes \cite{Burkov2011, Breitkreiz2019}. 
 
The solution of \eqref{eq2s} together with the solution 
 for $\mu_s$ from \eqref{muss} read
\begin{subequations}
\begin{align}
\mu_n =& -\theta \frac{\varepsilon_c}{k_0} \frac{x^2}{W}\, E +z \, E, 
\label{phrs} \\
\mu_a=& -\frac{\theta}{k_0} WE\Bigg[2\bigg(\frac{l_c}{W}\bigg)^2
\bigg(\varepsilon_F+\varepsilon_c\gamma\bigg)
\bigg(1 
 -\frac{\cosh\frac{x}{l_c}}{\frac{l_{ns}}{\theta l_c}\sinh\frac{W}{2l_c} +\cosh\frac{W}{2l_c}}\bigg)
+\varepsilon_c\gamma\frac{x^2}{W^2}\Bigg],\\
\mu_{s\pm} =& \; \mu_n|_{x=W/2} -\frac{\bar{nv}_{s}^z}{n_sv }l_{ns}E
= -\frac{\varepsilon_c}{k_0}\theta 
\Big(\frac{1}{4}+\frac{l_{ns}}{\theta W}
\Big)W\, E+z \, E.
\end{align}
\label{sols}
\end{subequations}

\subsection{Hollow-cylinder geometry}

We now calculate the 
potential $\mu_n$ in the geometry of a hollow cylinder.
The derivation of the Boltzmann equation \eqref{be2} and summation over states
leading to Eqs.\ \eqref{eqss} is unmodified. 
For the calculation of $\mu_n$ it is sufficient to 
consider  Eqs.\ \eqref{jnpms}, \eqref{eqss},
and \eqref{jcpms}, summed over $\pm$ and
the sum of $n$  and $c$ states, which then gives
\begin{subequations}
\begin{align}
\vj_n &= - n_nD\vn \mu_n, \\
\vn\cdot(\vj_n+\vj_c) &= - s(r)\vn\cdot\vj_s ,\\
\vn\cdot\vj_c &= -2\frac{\varepsilon_c}{k_0}\frac{n_s v}{r_o-r_i}\, E,\\
\vn\cdot\vj_s & = \frac{\varepsilon_c}{k_0} n_s v\, E,
\end{align}
\label{cyls}
\end{subequations}
where we used \eqref{jcpms}, $n_n \gg n_c$, 
the cancellation of the total equilibrium current,
\begin{align}
\pi (r_o^2-r_i^2) \sum_\pm \bar{nv}_{c\pm}  
+ 2\pi(r_o+r_i)\bar{nv}^z_s = 0,
\end{align}
and the not altered result for the 2D density-of-states 
weighted integral over velocity of surface states \eqref{nvs}.

 In cylinder coordinates the divergence  $\vn\cdot\vj_n$ 
 becomes $(\partial_r+1/r)j_n^r$, leading to the 
 differential equation 
 \begin{equation}
 (\partial_r+1/r)\partial_r \mu_n = -2\theta\frac{\varepsilon_c}{k_0}
 \frac{E}{r_o-r_i}.
 \end{equation}
The solution satisfying the boundary conditions given 
in \eqref{cyls} reads
\begin{equation}
\mu_n = -\frac{n_s v}{n_nD} \frac{\varepsilon_c}{k_0}
\bigg( \frac{r^2/2}{r_o-r_i} 
-\frac{r_or_i}{r_o-r_i}\ln\frac{r}{r_i}\bigg)E +zE.
\end{equation}
Defining the 
Hall angle for the PHE as $
\theta_\mathrm{PHE} \equiv [\mu_n(r_o)-
\mu_n(r_i)]/(r_o-r_i)E$ in the limit $r_o\gg r_i$,  
we obtain 
\begin{equation}
\theta_\mathrm{PHE} = -\frac{n_s v}{n_nD}\frac{\varepsilon_c}{2 k_0}.\label{phe}
\end{equation}

\clearpage

\subsection{Current density and conductivity}
To obtain the longitudinal resistivity, we calculate
the current densities in the $z$ direction.
From  \eqref{jnpms} we obtain 
\begin{align}
j^z_n = -n_nD\, E.
\end{align}
The non-equilibrium current contribution of chiral bulk particles and surface
states is given by 
\begin{align}
j^z_{s/c}(x) &= \frac{1}{W}\int \frac{dk_y dk_zd\omega}{(2\pi)^2} A(\vk, \vx,\omega) n_F'(\varepsilon_\vk)\mu(\vk, \vx) v^z_\vk
\nonumber \\
\Rightarrow\;\;\;  j^z_s(x) &= \delta(x\pm W/2)\;\bar{nv}^z_{s}\,\mu_{s\pm}
 \label{eqjs}\\
\Rightarrow\;\;\;  j^z_c(x)  &= \sum_\pm
\bar{nv}^z_{c\pm}\,\mu_{n\pm}, \label{eqjc}
\end{align}
where we used that the non-equilibrium part of the occupation
function reads $n_F'(\varepsilon_\vk)\mu(\vk, \vx) $
(the equilibrium parts cancel each other as discussed in 
the main text) and 
neglected the variation of $\mu(\vk, \vx)$ with $\vk$ which 
would give a correction of order $1/W$.
Now using Eq.\ \eqref{muna} and 
the solutions \eqref{sols} we obtain
\begin{align}
\frac{j^z_c(x)}{j^z_n} ={} & - \theta^2\frac{\varepsilon_c^2}{k_0^2}\Bigg[2 \Big(\frac{x}{W}\Big)^2 
+ \Big(\frac{2 l_c}{W}\Big)^2\frac{\varepsilon_F^2}{\varepsilon_c^2}\bigg(
\frac{\cosh\frac{x}{l_c}}{\tfrac{ l_{ns}}{\theta l_c}\sinh\frac{W}{2l_c}+ \cosh\frac{W}{2l_c}}-1 \bigg)\Bigg],\\
\frac{j^z_s(x)}{j^z_n}={} & \theta^2 \frac{\varepsilon_c^2}{k_0^2}\bigg(\frac{1}{4}+\frac{ l_{ns}}{\theta W}\bigg)
W\,\sum_\pm \delta(x\pm W/2) .
\end{align}
The resistivity is given by 
\begin{equation}
\sigma^{zz} = \frac{\bar{j^z_c}+\bar{j^z_s}+j^z_n}{E}, \;\;\;\;\;\;\;
\bar{j^z_i} = \frac{1}{W}\int dx\, j^z_i(x).
\end{equation}
In the absence of the PHE, i.e., if the potential would be  homogeneous in the $x$ direction, the contributions of 
surface und chiral bulk states would cancel each other
(up to the neglected correction of order $1/W$). In this case, the current would be carried mainly via the normal bulk states
and the resistivity would 
be given   by $\sigma^{zz}_0= j^z_n/E = -n_nD$, which 
is also what one would obtain from the Drude formula for 
the conductivity of the infinite system. Now 
taking into account the effect of the PHE, the full 
conductivity 
in terms of $\sigma^{zz}_0$ becomes
\begin{align}
\frac{\sigma^{zz}}{\sigma^{zz}_0}=&\; 1+\frac{4}{3}\theta^2_\mathrm{PHE} \bigg(1+\frac{6l_{ns}}{W\theta}\bigg)
\bigg(1+\frac{\varepsilon_F^2}{\varepsilon_c^2}\xi\bigg),
\label{ress}\\
&\xi = \frac{\frac{2 l_c}{W}-\frac{\big(\frac{2 l_c}{W}\big)^2}{\frac{ l_{ns}}{\theta l_c}+\coth\frac{W}{2l_c}}}{\frac{W}{6 l_c}+\frac{l_{ns}}{\theta l_c}}
=
\begin{cases} 
0 & l_c \ll W \\
1 & l_c \gg W,\;  l_{ns}/\theta.
\end{cases}
\end{align}

\end{document}